\documentclass[a4paper,11pt]{article}
\pdfoutput=1

\usepackage{jcappub} 
\usepackage{natbib}
\setcitestyle{square,comma,numbers,sort&compress}
\usepackage{enumerate}
\usepackage{tabularx}
\usepackage[left = 1.0in, top=1.0in,right=1.0in,bottom=1.0in,a4paper]{geometry}
\usepackage[dvipsnames]{xcolor}
\usepackage{comment}
\usepackage[caption=false]{subfig}
\usepackage{cancel}

\usepackage[section]{placeins}
\usepackage{listings}
\usepackage{amsbsy}
\newcolumntype{Y}{>{\centering\arraybackslash}X}

\definecolor{myRED}{rgb}{0.8, 0.25, 0.33}
\usepackage[normalem]{ulem}
\usepackage{dsfont}
\usepackage{pbox}
\usepackage{graphicx}
\usepackage{multirow}
\usepackage{tikz}
\usetikzlibrary{arrows}
\usepackage{enumitem} 
\usepackage{ulem}
\usepackage{soul}
\usepackage{lipsum}

\title{\boldmath\huge  Explaining PTA Results by Metastable Cosmic Strings from SO(10) GUT}
\author{Stefan Antusch,}
\author{Kevin Hinze,}
\author{Shaikh Saad}

\affiliation{Department of Physics, University of Basel, Klingelbergstrasse\ 82, CH-4056 Basel, \\Switzerland}

\emailAdd{stefan.antusch@unibas.ch, kevin.hinze@unibas.ch, shaikh.saad@unibas.ch}
\abstract{
In a recent paper  ({\hypersetup{urlcolor=black}\href{https://doi.org/10.1103/PhysRevD.108.095053}{\textit{Phys.Rev.D 108 (2023) 9, 095053}}}), we have demonstrated that the 2023 PTA results, which hint at a stochastic gravitational wave (GW) background at nanohertz frequencies, point towards a promising model-building route for realizing $SO(10)$ Grand Unification with embedded inflation. The proposed supersymmetric scenario solves the doublet-triplet splitting without fine-tuning, accounts for charged fermion and neutrino masses, avoids conflicts with current proton decay bounds, and includes only representations no larger than the adjoint. It features multi-step breaking of $SO(10)$ to the Standard Model gauge symmetry, with inflation embedded such that metastable cosmic strings are produced at the end of inflation. This cosmic string network generates a stochastic GW background that can explain the PTA results. In this paper, we provide a detailed analysis of the singled out GUT model class, focusing on how the gauge coupling unification condition affects the scales of multi-step $SO(10)$ breaking and the preferred GW spectra. The lowest breaking scale, linked to inflation, the generation of right-handed neutrino masses for the seesaw mechanism, and metastable cosmic string production, coincides with the range suggested by the PTA results.
}
\makeatletter
\gdef\@fpheader{}
\makeatother
\begin{document}
\maketitle
\flushbottom

\section{Introduction}
Recently, pulsar timing arrays (PTAs), CPTA~\cite{Xu:2023wog}, EPTA~\cite{Antoniadis:2023ott}, NANOGrav~\cite{NANOGrav:2023gor}, and PPTA~\cite{Reardon:2023gzh} (see also Ref.~\cite{InternationalPulsarTimingArray:2023mzf}) have found strong evidence for Hellings-Downs angular correlation~\cite{Hellings:1983fr} between the time-of-arrival perturbations of multiple pulsars. This observation of the Hellings-Downs pattern -- the smoking-gun signal of a stochastic gravitational wave background (SGWB) -- points towards the GW origin of this signal at nanohertz frequencies. Remarkably, the GW spectrum of a metastable cosmic string~\cite{Kibble:1976sj,Vilenkin:1982hm}  network at such low frequencies fits very well~\cite{NANOGrav:2023hvm,Antoniadis:2023zhi} with the recent PTA data. This has generated some excitement in the particle physics community, cf.\  Refs.~\cite{Antusch:2023zjk,Buchmuller:2023aus,Fu:2023mdu,
Lazarides:2023rqf,Ahmed:2023rky,Afzal:2023cyp,Maji:2023fhv,Ahmed:2023pjl,Afzal:2023kqs,King:2023wkm,Roshan:2024qnv,Ahmed:2024iyd,Antusch:2024ypp,Maji:2024pll}\footnote{ Note however that Ref.~\cite{Fu:2023mdu} considered a symmetry breaking chain, which does not lead to metastable cosmic strings. A similar observation was also made in Ref.~\cite{Lazarides:2023rqf}. Ref.~\cite{Fu:2023mdu} utilized $126_H$ in the Higgs sector, consequently, it leaves a discrete $Z_2$ symmetry unbroken, which leads to stable $Z_2$  strings~\cite{Kibble:1982ae}. } for works on metastable strings, and Refs.~\cite{Lazarides:2023ksx,Yamada:2023thl,Servant:2023mwt,
Lazarides:2023bjd,Pallis:2024mip} for other works on unstable cosmic strings. 
Whether the signal indeed originates from a cosmic string network can be fully tested in multiple GW observatories that include the currently operating detectors LIGO-Virgo-KAGRA (LVK)~\cite{LIGOScientific:2014pky,VIRGO:2014yos,KAGRA:2018plz}, as well as upcoming experiments  such as Laser Interferometer Space Antenna (LISA)~\cite{Audley:2017drz}, Big Bang Observer (BBO)~\cite{Corbin:2005ny}, DECi hertz Interferometer Gravitational wave Observatory (DECIGO)~\cite{Seto:2001qf}), Einstein Telescope (ET)~\cite{Sathyaprakash:2012jk}, and Cosmic Explorer (CE)~\cite{Evans:2016mbw}.

As recently pointed out in Ref.~\cite{Antusch:2023zjk}, metastable cosmic strings typically emerge within a particular class  of promising $SO(10)$ Grand Unified Theories (GUTs)~\cite{Pati:1973rp,Pati:1974yy, Georgi:1974sy, Georgi:1974yf, Georgi:1974my, Fritzsch:1974nn}. This class of
models is supersymmetric (SUSY) and, therefore, offers a solution to, or at least ameliorates, the hierarchy problem. SUSY also provides flat directions, enabling us to implement cosmic inflation~\cite{Guth:1980zm, Albrecht:1982wi, Linde:1981mu, Linde:1983gd}, which addresses the horizon and flatness problems of the standard Big Bang cosmology, and provides the seed for structure formation in the observable universe. These promising models utilize small-dimensional representations, avoiding the unified coupling  becoming non-perturbative below the Planck scale.  In SUSY GUTs, color triplet Higgses, which are partners of the electroweak Higgs doublets, typically need to be superheavy as they induce potentially dangerous $d=5$ operators~\cite{Sakai:1981pk,Weinberg:1981wj}. This introduces the so-called  doublet–triplet splitting (DTS) problem~\cite{Randall:1995sh,Yamashita:2011an}. Within this framework, utilizing the Dimopoulos-Wilczek  mechanism~\cite{Dimopoulos:1981xm,Srednicki:1982aj}, the DTS problem is solved without fine-tuning (see also Refs.~\cite{Babu:1993we,Babu:1994kb,Berezhiani:1996bv,Barr:1997hq,Chacko:1998jz,Babu:1998wi,Babu:2002fsa,Kyae:2005vg,Babu:2010ej,Wan:2022glq}) in a way that simultaneously keeps dangerous dimension-five nucleon decay~\cite{Sakai:1981pk,Weinberg:1981wj} under control. Additionally, the phase transition scale leading to the formation of a metastable cosmic string network is identified with the scale of inflation as well as the $B-L$ breaking scale, leading to non-zero neutrino masses via the type-I seesaw mechanism~\cite{Minkowski:1977sc,Yanagida:1979as,Glashow:1979nm,Gell-Mann:1979vob,Mohapatra:1979ia,Schechter:1980gr,Schechter:1981cv}. This GUT setup is highly attractive since it solves multiple challenges of the Standard Model (SM) of particle physics and cosmology.

In this work, we provide a more detailed analysis of the $SO(10)$ GUT scenario with embedded inflation that has been ``singled out'' in Ref.~\cite{Antusch:2023zjk} and which leads to the formation of a metastable cosmic string network. The scenario features lower-dimensional fields, no higher than the adjoint representation. A special feature of the model class is the presence of two copies of adjoint Higgs superfields. We identify the symmetry breaking chains that accommodate metastable strings, construct the superpotential, and compute the full mass spectrum.   We also scrutinize the Yukawa part of the superpotential and show the consistency of charged fermion and neutrino masses and mixings. Since in our scenario several states acquire masses somewhat below the GUT scale, they significantly alter gauge coupling unification compared to the Minimal Supersymmetric Standard Model (MSSM). We therefore perform a dedicated analysis of gauge coupling unification and proton decay constraints, demonstrating their compatibility while obtaining the common scale that signifies the scale of inflation, neutrino mass generation, and metastable cosmic string network formation. Furthermore, we compute the GW spectrum of the string network arising from our scenario and briefly discuss the implications of some amount of entropy production, e.g.\ from a late-decaying modulus field -- often present in SUSY theories -- inducing a temporary matter domination phase in the evolution of the universe, which may somewhat dilute the GW spectrum.

This paper is organized in the following way. In Sec.~\ref{sec:string}, we discuss how metastable cosmic strings appear within the proposed framework. All model details, including the full superpotential, derivation of the mass matrices, embedding of inflation, and fermion mass generation are worked out in Sec.~\ref{sec:model}. Implications of the model class, in particular for gauge coupling unification, proton decay, and the GW spectrum from metastable cosmic strings, are studied in Secs.~\ref{sec:unification} and~\ref{sec:spectrum}. Finally, we conclude in Sec.~\ref{sec:conclusion}.

\section{Metastable Cosmic Strings from \textit{SO}(10) GUT}\label{sec:string}
\subsection{Topological Defects}\label{sec:2point1}
The idea of Grand Unification is that the SM gauge group $\mathcal G_{321}\equiv SU(3)_C\times SU(2)_L\times U(1)_Y$ is contained within a larger group at a high energies. Subsequent symmetry breaking phase transitions eventually lead to the SM at low energies. Based on the symmetry breaking pattern, these transitions may generate diverse types of topological defects~\cite{Vilenkin:2000jqa}. The most common topological defects are monopoles~\cite{Kibble:1976sj,Preskill:1979zi}, cosmic strings~\cite{Kibble:1976sj}, and domain walls~\cite{Weinberg:1974hy}. If a symmetry group $G$ is broken to a subgroup $G^\prime$, i.e., $G\to G^\prime$, the type of defect that arises depends on the topology (specifically, the homotopy group, cf. Ref.~\cite{Coleman:1975qj}) of the manifold of degenerate vacuum states. This manifold contains the quotient space $G/G^\prime$. If it is not simply connected, such that closed loops cannot be continuously shrunk to a
point, i.e.\ when its first homotopy group is non-trivial ($\pi_1(G/G^\prime)\neq {1}$), cosmic strings are formed. The simplest example that leads to the formation of cosmic strings is $U(1)\to \mathds{1}$. On the other hand, if there are non-shrinkable two-dimensional surfaces in the vacuum manifold, i.e.\ when its second homotopy group is non-trivial ($\pi_2(G/G^\prime)\neq {1}$),  monopoles appear. A simple example where monopoles appear is the breaking chain $SU(2)\to U(1)$.

Moreover, composite defects can form from  symmetry breaking patterns of the type $G\to G^\prime\to G^{\prime\prime}$. In such a scenario, different types of defects may arise if the corresponding homotopy groups of the vacuum manifolds $G/G^\prime$ and $G^\prime/G^{\prime\prime}$ are non-trivial. Nevertheless, the presence of stable defects in the final phase is governed by the topology of $G/G^{\prime\prime}$. If we consider a symmetry breaking chain $SU(2)\to U(1)\to \mathds{1}$, monopoles are formed at the first stage, whereas cosmic strings are generated  at the latter phase transition. Since these two defects originate from the same parent group, pairs of monopoles and
antimonopoles can spontaneously nucleate along the strings by quantum tunneling and make the strings decay -- leading to the formation of a ``metastable'' cosmic string network. This hybrid defect is unstable (since the homotopy group $\pi_{1}(SU(2)/\mathds{1})$ is trivial), and the lifetime of the metastable string network depends on the nucleation rate.

\subsection{Metastable Strings}
In this work, we are interested in metastable cosmic strings from $\mathcal G_{10}\equiv SO(10)$  GUTs. 
Since the SM group contains a $U(1)$ factor, regardless of the symmetry-breaking chain, the appearance of stable superheavy monopoles is unavoidable  (since $\pi_2(\mathcal G_{10}/\mathcal G_{321})\neq 1$). Therefore, cosmic inflation after the monopole production phase transition is necessary to dilute their over-abundance. However, in our scenario, cosmic strings emerge at the end of inflation and thus remain undiluted. As discussed in Ref.~\cite{Antusch:2023zjk}, among the three promising $SO(10)$ GUT model building routes, only the model class featuring two adjoint Higgs representations can give rise to metastable cosmic strings. Depending on the vacuum expectation values (VEVs) of the two adjoints ($45_H$ and $45^\prime_H$) and a pair of $16_H+\overline{16}_H$ Higgses, various symmetry breaking chains may arise, three of which can lead to a viable   metastable cosmic string network:\\  
(a) $\langle 45_H\rangle > \langle 45_H^\prime\rangle > \langle 16_H\rangle, \langle \overline{16}_H\rangle$:
\begin{align}
SO(10)
&\xrightarrow[45_H]{M_\mathrm{GUT}} 
SU(3)_{C}\times SU(2)_{L} \times SU(2)_{R}\times U(1)_{B-L} \equiv \mathcal{G}_{3221}
\nonumber\\
&\xrightarrow[45_H^\prime]{M_I} 
SU(3)_{C}\times SU(2)_{L} \times U(1)_{R}  \times U(1)_{B-L} \equiv \mathcal{G}_{3211}
\nonumber\\
&\xrightarrow[16_H+\overline{16}_H]{M_{II}} 
SU(3)_{C}\times SU(2)_{L} \times U(1)_{Y} \equiv \mathcal{G}_{321}\;,\label{SB:a}
\end{align}
(b) $\langle 45_H^\prime\rangle > \langle 45_H\rangle > \langle 16_H\rangle, \langle \overline{16}_H\rangle$:
\begin{align}
SO(10)
&\xrightarrow[45_H^\prime]{M_\mathrm{GUT}} 
SU(4)_{C}\times SU(2)_{L} \times U(1)_{R} \equiv \mathcal{G}_{421}
\nonumber\\
&\xrightarrow[45_H]{M_I} 
SU(3)_{C}\times SU(2)_{L} \times U(1)_{R}  \times U(1)_{B-L} 
\nonumber\\
&\xrightarrow[16_H+\overline{16}_H]{M_{II}} 
SU(3)_{C}\times SU(2)_{L} \times U(1)_{Y}  \;,\label{SB:b}
\end{align}
(c) $\langle 45_H\rangle = \langle 45_H^\prime\rangle > \langle 16_H\rangle, \langle \overline{16}_H\rangle$:
\begin{align}
SO(10)
&\xrightarrow[45_H+45_H^\prime]{M_\mathrm{GUT}} 
SU(3)_{C}\times SU(2)_{L} \times U(1)_{R}  \times U(1)_{B-L} 
\nonumber\\
&\xrightarrow[16_H+\overline{16}_H]{M_{I}} 
SU(3)_{C}\times SU(2)_{L} \times U(1)_{Y}  \;.\label{SB:c}
\end{align} 
$M_\mathrm{GUT}$ represents the scale at which the GUT symmetry is broken, which is identified with the mass of the heaviest gauge boson (these masses are provided later in the text). 
In all the above symmetry breaking chains, cosmic inflation is associated with the last stage\footnote{Since we employ $16_H+\overline{16}_H$ superfields instead of the large representations $126_H+\overline{126}_H$, the emergence of automatic R-parity within the $SO(10)$ group is no longer feasible. Nonetheless, a discrete symmetry, like a $\mathbb{Z}_2$ symmetry (matter-parity), can be easily imposed.}  of the phase transitions, where scalars within $16_H+\overline{16}_H$ superfields play the role of the waterfall fields.  

As discussed in Sec.~\ref{sec:2point1}, a symmetry breaking chain $SU(2)\to U(1)\to \mathds{1}$ leads to monopoles at the first stage, whereas cosmic strings are generated  at the latter phase transition.  Therefore,   in the case (a), the breaking $SU(2)_{R}\to U(1)_{R}$ forms `blue' monopoles (here, we follow the nomenclature adopted in Ref.~\cite{Lazarides:2019xai}) and the subsequent breaking $U(1)_{R}\times U(1)_{B-L}\to U(1)_Y$ creates cosmic strings.  If these symmetry breaking scales are close, i.e., $M_I\simeq M_{II}$, as aforementioned in Sec.~\ref{sec:2point1}, nucleation of monopole-antimonopole pairs causes the strings to decay, leading to the emergence of  metastable cosmic strings. 
To be specific, the appearance of $SU(2)_R\subset SO(10)$ in the first breaking chain is obtained by one of the adjoint VEVs $(15,1,1)\subset 45_H$ (here, for convenience, Pati-Salam decomposition is specified), whereas, the desired breaking  $SU(2)_R\to U(1)_R$ is achieved via the VEV of the second adjoint $(1,1,3)\subset 45_H^\prime$. For the breaking chain (b), the breaking $SU(4)_{C}\to 
SU(3)_{C} \times U(1)_{B-L}$ generates `red' monopoles, whereas cosmic strings are generated as before. In the third scenario, i.e.\ case (c), both the red and blue monopoles are present at the same scale and  nucleation of these monopoles prompt to metastable cosmic strings.
The last scenario, i.e.\ chain (c), can also lead to the appearance of stable monopoles~\cite{Lazarides:2024niy} (since $\pi_2(\mathcal{G}_{10}/ \mathcal{G}_{321})\neq 1$), however, in our scenario, not  at an observable level, as will be discussed later in the text. 

We note that instead of utilizing two copies of $45_H$ multiplets, one may wish to consider a scenario with one/two $54_H$ dimensional representation(s). However, a $54_H$ Higgs superfield can only break $SO(10)$ to the Pati-Salam group. Then, the $45_H$ ($45_H^\prime$) Higgs superfield would break Pati-Salam to the left-right (quark-lepton) symmetry group. To form metastable cosmic strings, one would further require $45_H^\prime$ ($45_H$) to provide an intermediate $\mathcal{G}_{3211}$ group before $16_H+\overline{16}_H$ acquire a VEV. Therefore, the setup considered in this work (consisting of two adjoints) is indeed the minimal choice.   

\section{Model Details}\label{sec:model}
In this section, we introduce the model and provide all necessary details concerning the symmetry breaking pattern and the mass spectrum. 
\subsection{Superpotential}
To achieve the desired symmetry breaking patterns of Eqs.~\eqref{SB:a}-\eqref{SB:c}, the proposed model class consists of only lower dimensional Higgs representations, namely $10_H\equiv H$, $16_H+\overline{16}_H\equiv \chi+\overline\chi$, and $45_H\equiv A$. The GUT symmetry is spontaneously broken down to the SM gauge group via the  VEVs of Higgs fields in the adjoint and spinorial representations. Finally, the SM symmetry breaking takes place at the electroweak (EW) scale when Higgses in the fundamental representations (and possibly in the spinorial representation) acquire VEVs. The decomposition of these supermultiplets under various subgroups of $SO(10)$) are summarized in Table~\ref{tab:decomposition}.

\begin{table}[t!]
\centering
\footnotesize
\resizebox{0.9\textwidth}{!}{
\begin{tabular}{|c|c|c|c|c|c|}
\hline

\textbf{$\mathcal{G}_{10}$}  &
\textbf{$\mathcal{G}_{422}$}  & \textbf{$\mathcal{G}_{421}$}  & \textbf{$\mathcal{G}_{3221}$\color{gray}$\not\subset \mathcal{G}_{421}$} & \textbf{$\mathcal{G}_{3211}$} & \textbf{$\mathcal{G}_{321}$}  \\ 

\hline \hline
\multirow{2}{*}{$10$}&
$(6,1,1)$ &  $(6,1,0)$ & $(3,1,1,-1/3)$ & $(3,1,0,-1/3)$ & \pbox{5cm}{\vspace{2pt} $(3,1,-1/3) \equiv H^{(3,1,-\frac{1}{3})}_{(6,1,1)}$ \vspace{2pt}} \\  
 &&&$(\overline 3,1,1,1/3)$ & $(\overline 3,1,0,1/3)$ & \pbox{5cm}{\vspace{2pt}$(\overline 3,1,1/3) \equiv H^{(\overline 3,1,\frac{1}{3})}_{(6,1,1)}$ \vspace{2pt}}\\ \cline{2-6}

&
$(1,2,2)$& $(1,2,1/2)$ &$(1,2,2,0)$ &$(1,2,1/2,0)$ & \pbox{5cm}{\vspace{2pt} $(1,2,1/2) \equiv H^{(1,2,\frac{1}{2})}_{(1,2,2)}$ \vspace{2pt}}\\
& &$(1,2,-1/2)$ &&$(1,2,-1/2,0)$& \pbox{5cm}{\vspace{2pt}$(1,2,-1/2) \equiv H^{(1,2,-\frac{1}{2})}_{(1,2,2)}$ \vspace{2pt}} \\  

\hline \hline
\multirow{2}{*}{$16$}&
 $(4,2,1)$ & $(4,2,0)$ & $(3,2,1,1/6)$ & $(3,2,0,1/6)$ & \pbox{5cm}{\vspace{2pt}$(3,2,1/6) \equiv\chi^{(3,2,\frac{1}{6})}_{(4,2,1)}$\vspace{2pt}}\\  
 &&& $(1,2,1,-1/2)$ &$(1,2,0,-1/2)$  &\pbox{5cm}{\vspace{2pt} $(1,2,-1/2) \equiv\chi^{(1,2,-\frac{1}{2})}_{(4,2,1)}$\vspace{2pt}}\\ \cline{2-6}

 &$(\overline 4,1,2)$ &\color{blue}$(\overline 4,1,1/2)$ & $(\overline 3,1,2,-1/6)$ &\color{blue}$(\overline 3,1,1/2,-1/6)$ &\pbox{5cm}{\vspace{2pt} $(\overline 3,1,1/3) \equiv\chi^{(\overline 3,1,\frac{1}{3})}_{(\overline 4,1,2)}$\vspace{2pt}}\\  
 &&\color{red}$(\overline 4,1,-1/2)$ & &\color{red} $(\overline 3,1,-1/2,-1/6)$ &\pbox{5cm}{\vspace{2pt} $(\overline 3,1,-2/3) \equiv\chi^{(\overline 3,1,-\frac{2}{3})}_{(\overline 4,1,2)}$\vspace{2pt}}\\
  && &$(1,1,2,1/2)$&\color{blue}$(1,1,1/2,1/2)$ &\pbox{5cm}{\vspace{2pt} $(1,1,1) \equiv\chi^{(1,1,1)}_{(\overline 4,1,2)}$\vspace{2pt}}\\  
 &&& &\color{red}$(1,1,-1/2,1/2)$&\pbox{5cm}{\vspace{2pt} $(1,1,0) \equiv\chi^{(1,1,0)}_{(\overline 4,1,2)}$\vspace{2pt}}\\ 
 
\hline \hline
\multirow{2}{*}{$45$}&
$(1,1,3)$ & $(1,1,1)$ & $(1,1,3,0)$ & $(1,1,1,0)$ &\pbox{5cm}{\vspace{2pt} $(1,1,1) \equiv A^{(1,1,1)}_{(1,1,3)}$\vspace{2pt}}\\  
 && $(1,1,0)$ &&$(1,1,0,0)$  &\pbox{5cm}{\vspace{2pt} $(1,1,0) \equiv A^{(1,1,0)}_{(1,1,3)}$\vspace{2pt}}\\
 && $(1,1,-1)$ && $(1,1,-1,0)$ &\pbox{5cm}{\vspace{2pt}$(1,1,-1) \equiv A^{(1,1,-1)}_{(1,1,3)}$\vspace{2pt}}\\ \cline{2-6}

 &$(1,3,1)$&$(1,3,0)$&$(1,3,1,0)$ &$(1,3,0,0)$ &\pbox{5cm}{\vspace{2pt} $(1,3,0) \equiv A^{(1,3,0)}_{(1,3,1)}$\vspace{2pt}}\\ \cline{2-6}

 &$(6,2,2)$ &\color{blue}$(6,2,1/2)$ & $(3,2,2,-1/3)$ &\color{blue}$(3,2,1/2,-1/3)$ &\pbox{5cm}{\vspace{2pt} $(3,2,1/6) \equiv A^{(3,2,\frac{1}{6})}_{(6,2,2)}$\vspace{2pt}}\\  
 &&\color{red}$(6,2,-1/2)$ & &\color{red} $(3,2,-1/2,-1/3)$ &\pbox{5cm}{\vspace{2pt}$(3,2,-5/6) \equiv A^{(3,2,-\frac{5}{6})}_{(6,2,2)}$\vspace{2pt}}\\
  && &$(\overline 3,2,2,1/3)$&\color{blue}$(\overline 3,2,1/2,1/3)$ &\pbox{5cm}{\vspace{2pt} $(\overline 3,2,5/6) \equiv A^{(\overline 3,2,\frac{5}{6})}_{(6,2,2)}$\vspace{2pt}}\\  
 &&& &\color{red}$(\overline 3,2,-1/2,1/3)$&\pbox{5cm}{\vspace{2pt} $(\overline 3,2,-1/6) \equiv A^{(\overline 3,2,-\frac{1}{6})}_{(6,2,2)}$\vspace{2pt}}\\ \cline{2-6}

 &$(15,1,1)$ &$(15,1,0)$ & $(1,1,1,0)$ &$(1,1,0,0)$& \pbox{5cm}{\vspace{2pt}$(1,1,0)\equiv A^{(1,1,0)}_{(15,1,1)}$\vspace{2pt}}\\  
 &&&$(3,1,1,2/3)$ &$(3,1,0,2/3)$&\pbox{5cm}{\vspace{2pt} $(3,1,2/3)\equiv A^{(3,1,\frac{2}{3})}_{(15,1,1)}$\vspace{2pt}}\\
  &&&$(\overline 3,1,1,-2/3)$ & $(\overline 3,1,0,-2/3)$&\pbox{5cm}{\vspace{2pt}$(\overline 3,1,-2/3)\equiv A^{(\overline 3,1,-\frac{2}{3})}_{(15,1,1)}$\vspace{2pt}}\\  
 &&&$(8,1,1,0)$ &$(8,1,0,0)$&\pbox{5cm}{\vspace{2pt} $(8,1,0)\equiv A^{(8,1,0)}_{(15,1,1)}$\vspace{2pt}}\\ \hline
 
\end{tabular}
}
\caption{ Decompositions of 10-, 16-, and 45-dimensional representations under various subgroups.  Some of the multiplets of  $\mathcal{G}_{3211}$ are presented in color (red and blue) to explicitly show their embedding within $\mathcal{G}_{421}$.  }\label{tab:decomposition}
\end{table}

Note that a  superfield in the fundamental representation contains    weak doublets   and color triplets (see Table~\ref{tab:decomposition}). If the GUT symmetry is broken by an adjoint field having a VEV of the form 
\begin{align}
\langle A\rangle \propto i\tau_2\otimes \mathrm{diag}(a_1,a_2,a_3,a_4,a_5),   
\end{align}
with generic $a_1=a_2=a_3$ and $a_4=a_5$, 
then both the weak doublets and the color triplets  acquire masses of the order of the GUT scale. One, therefore, in the absence of a specific mechanism requires a fine-tuning of the parameters to force the weak doublets to reside at the EW scale. To naturally obtain light doublets, we employ the ``Dimopoulos-Wilczek'' mechanism~\cite{Dimopoulos:1981xm,Srednicki:1982aj} (see also~\cite{Babu:1993we,Babu:1994kb,Berezhiani:1996bv,Barr:1997hq,Chacko:1998jz,Babu:1998wi,Babu:2002fsa,Kyae:2005vg,Babu:2010ej,Wan:2022glq,Antusch:2023zjk}). In this so-called missing VEV  mechanism, it is feasible to provide superheavy mass only to the color triplets, while the weak doublets remain light. This can be realized with  two possible   VEV directions:
\begin{itemize}
\item $a_4=a_5=0$, leading to $\langle A\rangle\propto B-L$   generator, 
\item $a_1=a_2=a_3=0$, leading to $\langle A\rangle\propto I_{3R}$   generator.
\end{itemize}

In the proposed model class, the entire  superpotential takes the following form: 
\begin{align}\label{eq:superpot}
W=    W_\mathrm{GUT-breaking}+  W_\mathrm{Inflation}+W_\mathrm{Mixed} 
+W_\mathrm{DTS}+W_\mathrm{Yukawa},
\end{align}
where the terms contained in $W_\mathrm{GUT-breaking}+  W_\mathrm{Inflation}+W_\mathrm{Mixed}$ provide a consistent symmetry breaking of the GUT group down to the SM gauge group, $W_\mathrm{Yukawa}$ describes the Yukawa interactions, and $W_\mathrm{DTS}$ is the part of the superpotential responsible for realizing DTS without fine-tuning.

\textbf{GUT symmetry breaking:}
First, let us consider the GUT symmetry breaking. The corresponding superpotential is composed of the following terms: 
\begin{align}
W_\mathrm{GUT-breaking}&\supset \frac{m_{45}}{2} \text{Tr}[45_H^2]+\frac{\lambda}{4\Lambda} \text{Tr}[45_H^4] +\frac{m_{45^\prime}}{2} \text{Tr}[45_H^{\prime 2}]+\frac{\lambda^\prime}{4\Lambda} \text{Tr}[45_H^{\prime 4}]  
\nonumber\\
&+\frac{\eta}{\Lambda} \text{Tr}[45_H45_H45_H^\prime 45_H^\prime]  , \label{eq:GUTbreaking}   
\end{align}
where the 45-plets acquire VEVs of the form
\begin{align}
\langle 45_H\rangle\propto i\tau_2\otimes \mathrm{diag}(a,a,a,0,0), \;\;\;\mathrm{and}\;\;\;
\langle 45_H^\prime\rangle\propto i\tau_2\otimes \mathrm{diag}(0,0,0,b,b).     \label{eq:DW}
\end{align}
These VEV alignments can be realized by the form of the  superpotential given in Eq.~\eqref{eq:GUTbreaking}. Here $\Lambda \gtrsim 10 M_\mathrm{GUT}$ is a cutoff scale. The unmixed terms with only $45_H$ and $45_H^\prime$ break $\mathcal G_{10}$ to the left-right symmetry group  $\mathcal G_{3221}$  and the quark-lepton symmetry group $\mathcal G_{421}$, respectively. The former breaks the generators in $(3,2,+1/6)+(3,2,-5/6)+(3,1,2/3)+c.c.$, whereas the latter breaks the generators in $(3,2,+1/6)+(3,2,-5/6)+(1,1,+1)+c.c.$. As a result, there would be additional massless states. These states obtain their masses from the mixed term given in Eq.~\eqref{eq:GUTbreaking}.

\textbf{Inflation:} The symmetry breaking achieved by these adjoint superfields leads to the production of superheavy monopoles, which need to be inflated away. To achieve this, one can e.g.\ employ supersymmetric hybrid inflation~\cite{Linde:1993cn,linde1991axions,Dvali:1994ms} at the last intermediate symmetry breaking stage (alternatively, one may utilize  tribrid inflation \cite{Antusch:2004hd,Antusch:2009vg,Antusch:2010va,Antusch:2024qpb}). In this setup, a GUT singlet superfield $S$ plays the role of the inflaton, while the waterfall fields $16_H+\overline{16}_H$ are responsible for breaking the remaining symmetry to the SM group and terminate inflation. Successful implementation of hybrid inflation is obtained via the following part of the superpotential:  
\begin{align}
W_\mathrm{Inflation} \supset \varkappa S(\overline{16}_H16_H-m_{16}^2) .  \label{eq:inflation}  
\end{align}
The scalar potential, $V_\mathrm{inf}$, arsing from $W_\mathrm{Inflation}$, is flat along the direction of 16-plets and provides a constant energy density $V_\mathrm{inf}=\varkappa^2 m_{16}^4$ that drives inflation. During inflation in this flat valley the VEV of $S$ generates masses for the 16-plets which prevents them from getting their VEVs. Although the potential from $W_\mathrm{Inflation}$ is completely flat at tree level, e.g.\ loop corrections or K\"ahler contributions in supergravity can generate the necessary tilt to realize successful inflation consistent with Cosmic Microwave
Background (CMB) observations (cf.\ \cite{Senoguz:2004vu,Bastero-Gil:2006zpr,Rehman:2009nq,Nakayama:2010xf,Antusch:2012jc,Buchmuller:2014epa,Schmitz:2018nhb}). Due to this rolling, the value of the inflation field decreases and beyond a critical value the 16-plets become tachyonic, which ends inflation and causes the ``waterfall'' phase transition where the 16-plets obtain their VEVs. Within our framework with embedded hybrid inflation, cosmic strings are produced only after the end of the inflation, hence they do not participate in inflation. It might be possible to partially inflate the strings, which, however, requires a different inflationary setup (cf.~\cite{Lazarides:2023rqf}), a scenario not considered in this work.

Since the phase transitions associated with the Higgs superfields $45_H^{(\prime)}$ have already taken place before inflation, produced monopoles are diluted away by the inflationary phase. Cosmic strings, on the other hand, remain undiluted.  Inflationary models suggest that in addition to scalar modes, tensor modes are also generated. Since the amplitude of a power spectrum is proportional to the energy scale of inflation, observations directly constrain this scale. The Planck~\cite{Planck:2018jri} and BICEP2-Keck~\cite{BICEP:2021xfz} experiments provide a stringent bound on the tensor-to-scalar ratio, $r<0.036$, which implies that $V_\mathrm{inf}\lesssim 10^{16}$ GeV. Therefore, a larger cosmic string formation scale  $m_{16}\gtrsim 10^{16}$ GeV corresponds to a smaller $\varkappa$ value to be compatible with this constraint.

\textbf{Mixed terms in the superpotential:} Note, however, that adjoint and  spinorial representations have component superfields that share the same SM quantum numbers, namely
\begin{align}
45_H^{(\prime)}, 16_H, \overline{16}_H\supset (1,1,1)+ (3,2,1/6)+ (\overline 3,1,-2/3) +c.c.\;. \end{align}
Therefore, to avoid additional would-be Goldstone bosons that  would spoil gauge coupling unification, these fields must have non-trivial mixed terms. These mixed terms are also essential for providing masses to the rest of the sub-multiplets of $16_H+\overline{16}_H$, since $W_\mathrm{Inflation}$ only gives mass to the singlet components.  However, a mixed term of the form, $\overline{16}_H 45_H 16_H$,  is not welcome since it would destabilize the VEV of the adjoints from the desired ``Dimopoulos-Wilczek forms'' of Eq.~\eqref{eq:DW}. This, consequently,  requires the introduction of additional two pairs of spinorial representations, $16_H^\prime+\overline{16}_H^\prime$ and $16_H^{\prime\prime}+\overline{16}_H^{\prime\prime}$, and we consider the following mixed terms:
\begin{align}
W_\mathrm{Mixed}&\supset
\overline{16}_H(\lambda_1 45_H+\lambda_1^\prime 1_H)16_H^\prime +\overline{16}_H^\prime (\lambda_245_H+\lambda_2^\prime 1_H^\prime)16_H
\nonumber\\&
+
\overline{16}_H(\lambda_3 45_H^\prime+\lambda_3^\prime 1_H^{\prime\prime})16_H^{\prime\prime} +\overline{16}_H^{\prime\prime} (\lambda_4 45_H^\prime+\lambda_4^\prime 1_H^{\prime\prime\prime})16_H \;. \label{eq:mixed}
\end{align}
Here, we introduced the ``sliding singlets'' 
$1^{(\prime,\prime\prime,\prime\prime\prime)}_H$ that have VEVs determined by the above superpotential.  Moreover, the $F$-terms from $16_H+\overline{16}_H$ and $1_H^{(\prime,\prime\prime,\prime\prime\prime)}$  forbid the $16_H^\prime+\overline{16}_H^\prime$, $16_H^{\prime\prime}+\overline {16}_H^{\prime\prime}$ to obtain VEVs. 

In addition to avoiding extra light states that would spoil gauge coupling unification, the mixed terms also play a crucial role in arranging that at the end of inflation, the VEV of $16_H,\overline{16}_H$ breaks the intermediate symmetry group to the one of the SM: While the superpotential term $W_\mathrm{Inflation}$ of Eq.~\eqref{eq:inflation} determines the value of the moduli of the VEVs after inflation via $|\langle 16_H\overline{16}_H\rangle| = m^2_{16}$, when $m^2_{16}$ is chosen real w.l.o.g., $W_\mathrm{Mixed}$ provides different mass contributions to some of the scalar components of $16_H,\overline{16}_H$ during inflation. Combining these mass contributions with the universal mass contributions from $W_\mathrm{Inflation}$, the component that obtains the smallest squared mass combination from $W_\mathrm{Mixed}$ becomes tachyonic first and triggers the waterfall that ends inflation in exactly this direction. We have explicitly checked that by suitably choosing the superpotential couplings and the singlet VEVs (during the inflationary phase) in $W_\mathrm{Mixed}$, it is indeed possible to ensure that the right-handed neutrino direction of $16_H,\overline{16}_H$ becomes tachyonic first such that the desired SM vacuum is obtained.\footnote{We note that with the chosen form of $W_\mathrm{Inflation}$ and $W_\mathrm{Mixed}$, in addition to the scalar component of $S$, also the scalar components of the singlets $1_H$, $1_H^{\prime}$, $1_H^{\prime\prime}$ or $1_H^{\prime\prime\prime}$ (or a combination of all five fields)  may act as the inflaton, provided that e.g.\ loop corrections or K\"ahler contributions in supergravity generate suitable corrections to their tree-level flat potentials before symmetry breaking.}

\textbf{Doublet-triplet splitting:} Finally, in the proposed model class, doublet-triplet splitting is obtained by the following terms in the superpotential: 
\begin{align}
&W_\mathrm{DTS}\supset \gamma_1 10_H 45_H 10^\prime_H +\frac{\gamma_2}{\Lambda} 10^\prime_H 45^{\prime 2}_H10^\prime_H + M_{16}\overline{16}^{\prime\prime}_H16^\prime_H \;. \label{eq:DTS}
\end{align} 
Note that due to the antisymmetric property of the $45$-dimensional representation, two copies of fundamental representations are needed to write the first term. Since $\langle 45_H\rangle\propto i\tau_2\otimes \mathrm{diag}(a,a,a,0,0)$, this term provides GUT scale masses to only color triplets, but the doublets remain massless, i.e.,
\begin{align}
&10_H \langle 45_H\rangle 10_H^\prime\supset
\cancelto{0}{\overline 2_H 2_H^\prime} 
+\cancelto{0}{\overline 2_H^\prime 2_H}
+\overline 3_H 3_H^\prime 
+\overline 3_H^\prime 3_H\;. \label{eq:DTS01}
\end{align}
The above term, however, leads to two pairs of light doublets from $10_H$ and $10^\prime_H$, which would spoil  gauge coupling unification. To prevent this, we utilize the higher-dimensional term $10^\prime_H 45^{\prime 2}_H10^\prime_H$ that only gives rise to superheavy masses to the extra doublet pair from $10^\prime_H$, i.e.,
\begin{align}
&10_H^\prime 45_H^{\prime 2}10_H^\prime \supset
\overline 2_H^\prime  2_H^\prime 
+\cancelto{0}{\overline 3_H^\prime  3_H^\prime}\;. \label{eq:DTS03}
\end{align}
Despite the inclusion of the two aforementioned terms Eqs.~\eqref{eq:DTS01} and~\eqref{eq:DTS03}, one pair of color-triplets and an additional pair of weak doublets still remain massless. The remaining term $\overline{16}^{\prime\prime}_H16^\prime_H$ in the superpotential Eq.~\eqref{eq:DTS} ensures large masses to these states.

With the above superpotential, we have achieved DTS in a manner that forbids all leading order $d=5$ proton decay contributions. We note that instead of the higher-dimensional term $10^\prime_H 45^{\prime 2}_H10^\prime_H$ in Eq.~\eqref{eq:DTS}, one could introduce a  direct mass term $10^\prime_H 10^\prime_H$. However, the simultaneous presence of $10_H 45_H 10^\prime_H$ and   $10^\prime_H 10^\prime_H$ would induce an effective mass term $\overline 3_H 3_H$, reintroducing proton decay via color-triplet Higgses. Nevertheless, $d=5$ proton decay can be suppressed also in this case, since it depends on an effective triplet mass (which is not the physical mass), which can be superheavy (even $> 10^{19}$ GeV). A large effective triplet mass of this order would however imply an additional pair of doublet superfields as light as $\mathcal{O}(10^{12})$~GeV, potentially conflicting with gauge coupling unification. In the following, we do not pursue along this line and stick to the superpotential terms given in Eq.~\eqref{eq:DTS}. Therefore, instead of $p\to K^+\overline\nu$, which is typically the most dominant proton decay mode in supersymmetric GUTs, in our scenario the gauge boson mediated process $p\to \pi^0e^+$ is expected to be the leading proton decay channel.

\subsection{Mass Matrices}\label{sec:mass matrices}
Below, we present the mass matrices calculated from superpotential Eq.~\eqref{eq:superpot} (except the Yukawa sector, $W_\mathrm{Yukawa}$, which is considered in the next subsection). The columns ($c$) and rows ($r$) that define the basis are provided together with the mass matrices, using the notation introduced in Table~\ref{tab:decomposition}. We note that from the vanishing of the $F$-terms of the adjoints we have $m_{45}=\lambda \langle 45_H\rangle^2/(6\Lambda)$ and $m_{45^\prime}=\lambda^\prime \langle 45_H^\prime\rangle^2/(4\Lambda)$. 

\noindent \uline{Multiplet $(1,1,1)+c.c.$}: 
\begin{align}
r&: \chi^{(1,1,1)}_{(\overline 4,1,2)},\; \chi^{\prime (1,1,1)}_{(\overline 4,1,2)},\; \chi^{\prime \prime (1,1,1)}_{(\overline 4,1,2)},\; A^{(1,1,1)}_{(1,1,3)},\; A^{\prime (1,1,1)}_{(1,1,3)} \nonumber\\
c&: \overline\chi^{(1,1,-1)}_{(4,1,2)},\; \overline\chi^{\prime (1,1,-1)}_{(4,1,2)},\; \overline\chi^{\prime \prime (1,1,-1)}_{(4,1,2)},\; A^{(1,1,-1)}_{(1,1,3)},\; A^{\prime (1,1,-1)}_{(1,1,3)} \nonumber\\
\mathcal{M}_{(1,1,1)}&=    
\begin{pmatrix}
    0&0 & 4\lambda_4 \langle 45_H^\prime\rangle  & 0&0 
    \\
    0&0&M_{16}&2\lambda_1\langle\overline{16}_H\rangle &0
    \\
    4\lambda_3 \langle 45_H^\prime\rangle &0&0&0& 2\lambda_3 \langle \overline{16}_H\rangle
    \\
    0&2\lambda_2\langle 16_H\rangle&0& \frac{1}{2}m_{45}+ \frac{\eta \langle 45_H^\prime\rangle^2}{4\Lambda} &0
    \\
    0&0&2\lambda_4\langle 16_H\rangle&0&0
\end{pmatrix}. 
\label{eq:singlycharged}
\end{align}

\noindent \uline{Multiplet $(\overline 3,1,-2/3)+c.c.$}:
\begin{align}
r&: \chi^{(\overline 3,1,-\frac{2}{3})}_{(\overline 4,1,2)},\; \chi^{\prime (\overline 3,1,-\frac{2}{3})}_{(\overline 4,1,2)},\; \chi^{\prime \prime (\overline 3,1,-\frac{2}{3})}_{(\overline 4,1,2)},\; A^{(\overline 3,1,-\frac{2}{3})}_{(\overline 15,1,1)},\; A^{\prime (\overline 3,1,-\frac{2}{3})}_{(\overline 15,1,1)} \nonumber\\
c&: \overline\chi^{(3,1,\frac{2}{3})}_{(4,1,2)},\; \overline\chi^{\prime (3,1,\frac{2}{3})}_{(4,1,2)},\; \overline\chi^{\prime \prime (3,1,\frac{2}{3})}_{(4,1,2)},\; A^{(3,1,\frac{2}{3})}_{(15,1,1)},\; A^{\prime (3,1,\frac{2}{3})}_{(15,1,1)} \nonumber\\
\mathcal{M}_{(\overline 3,1,-\frac{2}{3})}&=    
\begin{pmatrix}
0&4\sqrt{\frac{2}{3}}\lambda_2\langle 45_H\rangle&0&0&0
\\
4\sqrt{\frac{2}{3}}\lambda_1\langle 45_H\rangle&0&M_{16}&2\lambda_1\langle \overline{16}_H\rangle&0
\\
0&0&0&0&2\lambda_3\langle \overline{16}_H\rangle
\\
0&2\lambda_2\langle 16_H\rangle&0&0&0
\\
0&0&2\lambda_4\langle 16_H\rangle&0&\frac{1}{2}m_{45_H^\prime}+\frac{\eta\langle 45_H\rangle^2}{6\Lambda}
\end{pmatrix}.
\label{eq:mass(3,1,-2/3)}
\end{align}

\noindent \uline{Multiplet $(3,2,1/6)+c.c.$}:
\begin{align}
r&: \chi^{(3,2,\frac{1}{6})}_{(4,2,1)},\; \chi^{\prime (3,2,\frac{1}{6})}_{(4,2,1)},\; \chi^{\prime \prime (3,2,\frac{1}{6})}_{(4,2,1)},\; A^{(3,2,\frac{1}{6})}_{(6,2,2)},\; A^{\prime (3,2,\frac{1}{6})}_{(6,2,2)} \nonumber\\
c&: \overline\chi^{(\overline 3,2,-\frac{1}{6})}_{(\overline 4,2,1)},\; \overline\chi^{\prime (\overline 3,2,-\frac{1}{6})}_{(\overline 4,2,1)},\; \overline\chi^{\prime \prime (\overline 3,2,-\frac{1}{6})}_{(\overline 4,2,1)},\; A^{(\overline 3,2,-\frac{1}{6})}_{(6,2,2)},\; A^{\prime (\overline 3,2,-\frac{1}{6})}_{(6,2,2)} \nonumber\\
\mathcal{M}_{(3,2,\frac{1}{6})}&=    
\begin{pmatrix}
0 & 2\sqrt{\frac{2}{3}} \lambda_2\langle 45_H\rangle & 2 \lambda_4\langle 45_H^\prime\rangle  & 0 & 0 
\\
2\sqrt{\frac{2}{3}}\lambda_1\langle 45_H\rangle & 0 &M_{16}& 2\lambda_1 \langle \overline{16}_H \rangle&0 
\\
2\lambda_3\langle 45_H^\prime \rangle&0&0&0&2\lambda_3 \langle \overline{16}_H\rangle
\\
0 & 2\lambda_2 \langle 16_H \rangle & 0& \frac{\eta\langle 45_H^\prime\rangle^2}{8\Lambda} & -\frac{\eta\langle 45_H\rangle \langle 45_H^\prime\rangle  }{4\sqrt{6}\Lambda}
\\
0&0&2\lambda_4 \langle 16_H\rangle&-\frac{\eta\langle 45_H\rangle \langle 45_H^\prime\rangle  }{4\sqrt{6}\Lambda}&\frac{\eta\langle 45_H\rangle^2 }{12\Lambda}
\end{pmatrix}.
\label{eq:mass(3,2,1/6)}
\end{align}

\noindent \uline{Multiplet $(1,2,-1/2)+c.c.$}:
\begin{align}
r&: H^{(1,2,-\frac{1}{2})}_{(1,2,2)},\; H^{\prime (1,2,-\frac{1}{2})}_{(1,2,2)},\; \chi^{(1,2,-\frac{1}{2})}_{(4,2,1)},\; \chi^{\prime (1,2,-\frac{1}{2})}_{(4,2,1)},\; \chi^{\prime \prime (1,2,-\frac{1}{2})}_{(4,2,1)} \nonumber\\
c&: H^{(1,2,\frac{1}{2})}_{(1,2,2)},\; H^{\prime (1,2,\frac{1}{2})}_{(1,2,2)},\;  \overline\chi^{(1,2,\frac{1}{2})}_{(\overline 4,2,1)},\; \overline\chi^{\prime (1,2,\frac{1}{2})}_{(\overline 4,2,1)},\; \overline\chi^{\prime \prime (1,2,\frac{1}{2})}_{(\overline 4,2,1)} \nonumber\\
\mathcal{M}_{(1,2,-\frac{1}{2})}&=    
\begin{pmatrix}
0&0&0&0&0
\\
0&-\frac{\gamma_2\langle 45_H^\prime\rangle^2}{4\Lambda} &0&0&0
\\
0&0&0&2\sqrt{6}\lambda_2 \langle 45_H\rangle& 2\lambda_4 \langle 45_H^\prime\rangle
\\
0&0&2\sqrt{6}\lambda_1 \langle 45_H \rangle&0&M_{16}
\\
0&0 & 2\lambda_3\langle 45_H^\prime\rangle & 0 &0 
\end{pmatrix}.  
\label{eq:doublet}
\end{align}

\noindent \uline{Multiplet $(\overline 3,1,1/3)+c.c.$}:
\begin{align}
r&: H^{(\overline 3,1,\frac{1}{3})}_{(6,1,1)}, H^{\prime (\overline 3,1,\frac{1}{3})}_{(6,1,1)}, \chi^{(\overline 3,1,\frac{1}{3})}_{(\overline 4,1,2)}, \chi^{\prime (\overline 3,1,\frac{1}{3})}_{(\overline 4,1,2)}, \chi^{\prime \prime (\overline 3,1,\frac{1}{3})}_{(\overline 4,1,2)} \nonumber\\
c&: H^{(3,1,-\frac{1}{3})}_{(6,1,1)}, H^{\prime (3,1,-\frac{1}{3})}_{(6,1,1)},  \overline\chi^{(3,1,-\frac{1}{3})}_{(4,1,2)}, \overline\chi^{\prime (3,1,-\frac{1}{3})}_{(4,1,2)}, \overline\chi^{\prime \prime (3,1,-\frac{1}{3})}_{(4,1,2)} \nonumber\\
\mathcal{M}_{(\overline 3,1,\frac{1}{3})}&=    
\begin{pmatrix}
0&\frac{\gamma_1\langle 45_H\rangle}{2\sqrt{6}}&0&0&0
\\
-\frac{\gamma_1\langle 45_H\rangle}{2\sqrt{6}}&0&0&0&0
\\
0&0&0&4\sqrt{\frac{2}{3}}\lambda_2 \langle 45_H\rangle& 4\lambda_4 \langle 45_H^\prime\rangle
\\
0&0&4\sqrt{\frac{2}{3}}\lambda_1 \langle 45_H\rangle&0&M_{16}
\\
0&0 & 4\lambda_3\langle 45_H^\prime\rangle & 0 &0 
\end{pmatrix}. 
\label{eq:triplet}
\end{align}

Furthermore, 
$A_{(15,1,1)}^{(8,1,0)}$ gets a mass $m_{45}$ and $A_{(15,1,1)}^{\prime(8,1,0)}$ has a mass   $\frac{1}{2}m_{45^\prime}+\eta \langle 45_H\rangle^2/(6\Lambda)$. While $A_{(1,3,1)}^{(1,3,0)}$ acquires a mass $\frac{1}{2}m_{45}+\eta \langle 45_H^\prime\rangle^2/(4\Lambda)$, $A_{(1,3,1)}^{\prime(1,3,0)}$ has a mass $m_{45^\prime}$. One linear combination  of $A_{(6,2,2)}^{(3,2,-5/6)}+c.c.$ and $A_{(6,2,2)}^{\prime(3,2,-5/6)}+c.c.$ is Goldstone, and the other set gets a mass $\eta(2\langle 45_H\rangle^2 + 3 \langle 45_H^\prime\rangle^2)/(24\Lambda)$. Also, the vector multiplets $X(3,2,-5/6)+c.c.$ and $Y(3,2,1/6)+c.c.$ get masses $M_{X}^2=g^2_\mathrm{GUT}(\langle 45_H\rangle^2/6+\langle 45_H^\prime\rangle^2/4)$ and $M_Y^2=M_X^2+4g^2_\mathrm{GUT}|\langle 16_H\rangle|^2$, respectively.

As can be seen from the doublet mass matrix, Eq.~\eqref{eq:doublet}, only a single pair remains light, while the color triplet mass matrix, Eq.~\eqref{eq:triplet}, does not contain any light states. The mass matrices derived in this section will play a crucial role in the gauge coupling unification analysis in Sec.~\ref{sec:unification}.

\subsection{Yukawa Sector}
In this section, we delineate the Yukawa interactions, incorporating the mass matrices and mixing angles of the SM fermions. In our setup, fermion masses are generated through the couplings of the matter superfields $16_F^i$  with the $10_H$ (where $i$ ranges from 1 to 3, representing the family index).    Since the only possible renormalizable term is inadequate to replicate the observed fermion mass spectrum, we include also the following higher-dimensional terms: 
\begin{align}
   W_\mathrm{Yukawa}&= Y_{10} 16_F 16_F 10_H  + \frac{Y_a}{\Lambda}(16_F 45_H)_{16}  (10_H 16_F)_{\overline{16}} + \frac{Y_b}{\Lambda}(16_F 45_H^\prime)_{16}  (10_H 16_F)_{\overline{16}}
    \nonumber\\ 
    &+ \frac{Y_{\nu^c}}{\Lambda}(\overline{16}_H 16_F)_1 (\overline{16}_H 16_F)_1\;.
\end{align}
In the above operators, we have suppressed the family indices. Moreover, $(16_F 45_H)_{16}$ represents the spinorial part obtained from the contraction of $16_F 45_H$, and so on. 

From the above Yukawa interactions, we obtain the following fermion mass matrices:
\begin{align}
&M_u= Y_{10} v^u_{10}-\sqrt{\frac{2}{3}}\frac{ \langle 45_H\rangle}{\Lambda}Y_a v^u_{10} +2\frac{ \langle 45_H^\prime\rangle}{\Lambda}Y_b v^u_{10}  \,,
\\
&M_d= Y_{10} v^d_{10}-\sqrt{\frac{2}{3}}\frac{ \langle 45_H\rangle}{\Lambda}Y_a v^d_{10} -2\frac{ \langle 45_H^\prime\rangle}{\Lambda}Y_b v^d_{10}  \,,
\\
&M_e= Y_{10} v^d_{10}+\sqrt{6}\frac{ \langle 45_H\rangle}{\Lambda}Y_a v^d_{10}  -2\frac{ \langle 45_H^\prime\rangle}{\Lambda}Y_b v^d_{10}  \,,
\\
&M_\nu^D=Y_{10}v^u_{10}+\sqrt{6}\frac{ \langle 45_H\rangle}{\Lambda}Y_a v^u_{10} +2\frac{ \langle 45_H^\prime\rangle}{\Lambda}Y_b v^u_{10} \,,
\\
&M_{\nu^c}= \frac{\langle \overline{16}_H\rangle^2}{\Lambda}Y_{\nu^c}\,. \label{righthanded}
\end{align}
Here, the EW scale VEVs are defined as $v^d_{10}\equiv \langle  H^{(1,2,\frac{1}{2})}_{(1,2,2)}\rangle$, and $v^u_{10}\equiv \langle H^{(1,2,-\frac{1}{2})}_{(1,2,2)}\rangle$. In family space, the $Y_{10}$ matrix is symmetric,  whereas $Y_a$ is anti-symmetric, and $Y_b$ is an arbitrary matrix. 

The heavy right-handed neutrino mass matrix is given by Eq.~\eqref{righthanded}, and the light (mainly) left-handed neutrinos acquire their masses via the type-I seesaw relation 
\begin{align}
\mathcal{M}_\nu= -  \left( M^D_\nu \right)^T  M_{\nu^c}^{-1} M^D_\nu \;. 
\end{align}
For Yukawa couplings set to unity, the neutrino mass scale is determined by
\begin{align}
m_\nu\sim  \frac{\Lambda v^2_\mathrm{ew}}{m^2_{16}},    
\end{align}
where $v_\mathrm{ew}$ denotes the electroweak scale.  The correct neutrino mass scale is then reproduced for $m_{16}\sim 10^{15-16}$ GeV. To be specific, assuming $\Lambda= 10 M_\mathrm{GUT}$, and setting the Yukawa couplings to unity, the correct neutrino mass scale is obtained for $m_{16}\sim M_\mathrm{GUT}\sim 10^{16}$ GeV. Instead, if the ``combination of Yukawas'' appearing in the seesaw formula is set to 0.01 (0.1), one  obtains the right neutrino mass scale with e.g.\  $m_{16}\sim 10^{15} (10^{16})$ GeV and $M_\mathrm{GUT}\sim 10^{16} (10^{17})$ GeV.  As will be shown in the following sections, this expected range of the seesaw scale is fully compatible with gauge coupling unification requirements and PTA observations.

Here, we briefly sketch how the baryon asymmetry of the universe could be generated within our model classes. When the inflaton decays into right-handed neutrinos and sneutrinos, this provides an attractive way of reheating the universe and for generating the baryon asymmetry via the well known leptogenesis mechanism~\cite{Fukugita:1986hr}. This decay of the inflaton field, in the scenario with embedded hybrid inflation,  proceeds via the mixture with the waterfall field. The waterfall field component then induces the decay to right-handed (s)neutrinos due to the non-renormalizable term that provides masses to the right-handed neutrinos. 
We note that an alternative scenario, where the inflation is directly identified as a right-handed sneutrino (cf.\ \cite{Antusch:2004hd,Antusch:2010va,Antusch:2010mv}), could realize non-thermal leptogenesis efficiently within tribrid inflation.

\section{Unification and Proton Decay}\label{sec:unification}

\textbf{Gauge coupling unification:} In the proposed model class, on top of the typical TeV scale MSSM states, there are additional multiplets that reside at intermediate mass scales.   Appearance of some of these states is due to the non-renormalizable terms in the superpotential and multistep $SO(10)$ breaking. Therefore, the unification of gauge couplings deviates from the automatic unification realized in the MSSM. Consequently, we perform a thorough investigation of gauge coupling unification in the following.

The renormalization group evolution of the gauge couplings $\alpha_i=g_i^2/(4\pi)$, at one-loop, is given by 
\begin{align}
    \alpha_i^{-1}(\mu)=\alpha_i^{-1}(\mu_0)+\frac{b_i}{2\pi}\ln\left(\frac{\mu}{\mu_0}\right)+\sum_J \frac{b_i^J}{2\pi}\ln\left(\frac{\mu}{M_J}\right)\mathcal{H}(\mu-M_J)\,,
\end{align}
where $b_i$ are the one-loop gauge coefficients of the respective group factor, whereas $b_i^J$ are the one-loop gauge coefficient of the multiplets $J$ with masses $M_J$. Moreover, $\mathcal{H}$ denotes the Heaviside-theta function, which is defined as $\mathcal{H}(\mu)=1$ for $\mu\geq0$ and $\mathcal{H}(\mu)=0$ otherwise. 

For the following discussion, we will fix the SUSY scale to $m_\mathrm{SUSY}=3\,$TeV. Also, we define $v_{45}=\langle 45_H\rangle$, $v_{45^\prime}=\langle 45_H^\prime\rangle$, and  $v_{16}=|\langle 16_H\rangle|=|\langle \overline{16}_H\rangle|$. Furthermore, we define $v_\mathrm{GUT}=\max\{v_{45},v_{45^\prime}\}$ and use the metastability condition $\min\{v_{45},v_{45^\prime}\} \approx   v_{16}$, where $v_{16}$ corresponds to the scale at which the  strings are produced.

We obtain the mass scales of the different chiral multiplets by approximately diagonalizing the mass matrices given in Eqs.~\eqref{eq:singlycharged}-\eqref{eq:triplet}. For simplicity of our analysis, we ignore the order one Clebsch–Gordan (CG) coefficients in the mass matrices and set all superpotential couplings to unity. Consequently, all the masses of the chiral superfields are roughly determined by three scales,  $v_\mathrm{GUT}$, $v_{16}$, and the cutoff scale, $\Lambda$.   Following these assumptions, we only vary the three scales $v_\mathrm{GUT}$, $v_{16}$ and $\Lambda$, while ensuring that $\Lambda/10\geq v_\mathrm{GUT}\geq v_{16}$.\footnote{Note that the additional free mass parameter $M_{16}$ does not effect the gauge coupling unification, but only the size of the unified gauge coupling $g_\mathrm{GUT}$. We therefore keep it fixed at $v_{16}$.} 

We note that ignoring the order one coefficients for the masses introduces some level of uncertainty. For example, if the mass of a color octet gets shifted by a factor of $e$ (Euler's number), this introduces a correction on $\alpha_3^{-1}(m_\mathrm{SUSY})$ of $3/(2\pi)$ corresponding to a $3\%$ shift of $g_3(m_\mathrm{SUSY})$. Since this can easily happen due to the ignored $\mathcal{O}(1)$ and CG factors, in our analysis, for the purpose of illustration, we accept points that realize gauge coupling unification within $3\%$ uncertainties on the low-energy measured values of each coupling $g_i$. 
We also note that within the precision we are working with, considering the renormalization group  analysis of the gauge couplings at the one-loop level appears sufficient.

\begin{table}[t!]
\centering
\renewcommand{\arraystretch}{0.85}
\begin{tabular}{|c|c|c|c|c|}
\hline

\textbf{$\mathcal{G}_{321}$} & \textbf{$\mathcal{G}_{3211}$}& \textbf{$\mathcal{G}_{3221}$}& \textbf{$\mathcal{G}_{10}$} & \textbf{$\mu$}  \\ 

\hline \hline
$H^{(1,2,\frac{1}{2})}_{(1,2,2)}$&$H^{(1,2,\frac{1}{2},0)}_{(1,2,2)}$& $H^{(1,2,2,0)}_{(1,2,2)}$&  $10_H$& $m_\mathrm{SUSY}$ \\ \hline 

-& $\chi^{(1,1,-\frac{1}{2},\frac{1}{2})}_{(\overline{4},1,2)}$& $\chi^{(1,1,2,\frac{1}{2})}_{(\overline{4},1,2)}$&  $16_H$& $v_{16}$ \\ \hline 

-& -& $A^{\prime(1,1,3,0)}_{(1,1,3)}$&  $45_H^\prime$& $v_{45^\prime}$ \\ \hline\hline

$A^{(8,1,0)}_{(15,1,1)}$& $A^{(8,1,0,0)}_{(15,1,1)}$& $A^{(8,1,1,0)}_{(15,1,1)}$&  $45_H$& $\frac{v^2_{\mathrm{GUT}}}{\Lambda}$ \\ \hline
$A^{\prime(8,1,0)}_{(15,1,1)}$& $A^{\prime(8,1,0,0)}_{(15,1,1)}$& $A^{\prime(8,1,1,0)}_{(15,1,1)}$&  $45^\prime_H$& $\frac{v^2_{\mathrm{GUT}}}{\Lambda}$ \\ \hline
$A^{(1,3,0)}_{(1,3,1)}$& $A^{(1,3,0,0)}_{(1,3,1)}$& $A^{(1,3,1,0)}_{(1,3,1)}$&  $45_H$& $\frac{v^2_{\mathrm{GUT}}}{\Lambda}$ \\ \hline
$A^{\prime(1,3,0)}_{(1,3,1)}$& $A^{\prime(1,3,0,0)}_{(1,3,1)}$& $A^{\prime(1,3,1,0)}_{(1,3,1)}$&  $45^\prime_H$& $\frac{v^2_{45^\prime}}{\Lambda}$ \\ \hline
$A^{\prime(3,2,-\frac{5}{6})}_{(6,2,2)}$ & $A^{\prime(3,2,-\frac{1}{2},-\frac{1}{3})}_{(6,2,2)}$ & $A^{\prime(3,2,2,-\frac{1}{3})}_{(6,2,2)}$  & $45^\prime_H$ & $\frac{v_\mathrm{GUT}^2}{\Lambda}$ \\ \hline

$A^{(1,1,1)}_{(1,1,3)}$& $A^{(1,1,1,0)}_{(1,1,3)}$& $A^{(1,1,3,0)}_{(1,1,3)}$&  $45_H$& max$\lbrace v_{16},\frac{v_\mathrm{GUT}^2}{\Lambda}\rbrace$ \\ \hline
$\chi^{\prime(1,1,1)}_{(\overline{4},1,2)}$& $\chi^{\prime(1,1,\frac{1}{2},\frac{1}{2})}_{(\overline{4},1,2)}$& $\chi^{\prime(1,1,2,\frac{1}{2})}_{(\overline{4},1,2)}$&  $16_H^\prime$& min$\lbrace v_{16},\frac{v_{16}^2 \Lambda}{v_\mathrm{GUT}^2}\rbrace$ \\ \hline
$\chi^{(1,1,1)}_{(\overline{4},1,2)}$& $\chi^{(1,1,\frac{1}{2},\frac{1}{2})}_{(\overline{4},1,2)}$& $\chi^{(1,1,2,\frac{1}{2})}_{(\overline{4},1,2)}$&  $16_H$& $v_{45^\prime}$ \\ \hline
$\chi^{\prime\prime(1,1,1)}_{(\overline{4},1,2)}$& $\chi^{\prime\prime(1,1,\frac{1}{2},\frac{1}{2})}_{(\overline{4},1,2)}$& $\chi^{\prime\prime(1,1,2,\frac{1}{2})}_{(\overline{4},1,2)}$&  $16_H^{\prime\prime}$& $v_{45^\prime}$ \\ \hline

$A^{\prime(\overline 3,1,-\frac{2}{3})}_{(15,1,0)}$& $A^{\prime(\overline 3,1,0,-\frac{2}{3})}_{(15,1,0)}$& $A^{\prime(\overline 3,1,1,-\frac{2}{3})}_{(15,1,1)}$&  $45_H^\prime$& max$\lbrace v_{16},\frac{v_\mathrm{GUT}^2}{\Lambda}\rbrace$ \\ \hline
$\chi^{\prime\prime(\overline 3,1,-\frac{2}{3})}_{(\overline{4},1,2)}$& $\chi^{\prime\prime(\overline 3,1,-\frac{1}{2},-\frac{1}{6})}_{(\overline{4},1,2)}$& $\chi^{\prime\prime(\overline 3,1,2,-\frac{1}{6})}_{(\overline{4},1,2)}$&  $16_H^{\prime\prime}$& min$\lbrace v_{16},\frac{v^2_{16}\Lambda}{v_\mathrm{GUT}^2}\rbrace$ \\ \hline

$A^{(3,2,\frac{1}{6})}_{(6,2,2)}$& $A^{(3,2,\frac{1}{2},-\frac{1}{3})}_{(6,2,2)}
$& $A^{(3,2,2,-\frac{1}{3})}_{(6,2,2)}$&  $45_H$& $v_{16}$ \\ \hline
$\chi^{\prime\prime(3,2,\frac{1}{6})}_{(4,2,1)}$& $\chi^{\prime\prime(3,2,0,\frac{1}{6})}_{(4,2,1)}$& $\chi^{\prime\prime(3,2,1,\frac{1}{6})}_{(4,2,1)}$&  $16^{\prime\prime}_H$& $v_{16}$ \\ \hline

$H^{\prime(1,2,\frac{1}{2})}_{(1,2,2)}$&$H^{\prime(1,2,\frac{1}{2},0)}_{(1,2,2)}$& $H^{\prime(1,2,2,0)}_{(1,2,2)}$&  $10_H^\prime$& $\frac{v^2_{45^\prime}}{\Lambda}$ \\ \hline 
$\chi^{\prime\prime(1,2,-\frac{1}{2})}_{(4,2,1)}$&$\chi^{\prime\prime(1,2,0,-\frac{1}{2})}_{(4,2,1)}$& $\chi^{\prime\prime(1,2,1,-\frac{1}{2})}_{(4,2,1)}$&  $16_H^{\prime\prime}$& $\frac{M_{16}v_{45^\prime}}{v_\mathrm{GUT}}$ \\ \hline

$\chi^{\prime\prime(\overline{3},1,\frac{1}{3})}_{(\overline{4},1,2)}$&$\chi^{\prime\prime(\overline{3},1,\frac{1}{2},-\frac{1}{6})}_{(\overline{4},1,2)}$& $\chi^{\prime\prime(\overline{3},1,2,-\frac{1}{6})}_{(\overline{4},1,2)}$ & $16_H^{\prime\prime}$& $\frac{M_{16}v_{45^\prime}}{v_\mathrm{GUT}}$ \\ \hline

\end{tabular}

\caption{ Approximate mass scales and mass eigenstates of the multiplets for gauge coupling unification for the scenario (a).  Only the states that reside below the GUT scale and relevant for unification are presented. The supermultiplets shown in the first three rows are relevant for symmetry breakings below the GUT scale. As described in the main text, we have fixed the SUSY scale at $m_\mathrm{SUSY}=3$ TeV.
}\label{tab:mixed-a-massscales}
\end{table}

\begin{table}[t!]
\centering
\renewcommand{\arraystretch}{0.85}
\begin{tabular}{|c|c|c|c|c|}
\hline

\textbf{$\mathcal{G}_{321}$} & \textbf{$\mathcal{G}_{3211}$}& \textbf{$\mathcal{G}_{421}$}& \textbf{$\mathcal{G}_{10}$} & \textbf{$\mu$}  \\ 

\hline \hline
$H^{(1,2,\frac{1}{2})}_{(1,2,2)}$&$H^{(1,2,\frac{1}{2},0)}_{(1,2,2)}$& $H^{(1,2,\frac{1}{2})}_{(1,2,2)}$&  $10_H$& $m_\mathrm{SUSY}$ \\ \hline 

-& $\chi^{(1,1,-\frac{1}{2},\frac{1}{2})}_{(\overline{4},1,2)}$& $\chi^{(\overline{4},1,-\frac{1}{2})}_{(\overline{4},1,2)}$&  $16_H$& $v_{16}$ \\ \hline 

-& -& $A^{(15,1,0)}_{(15,1,1)}$&  $45_H$& $v_{45}$ \\ \hline\hline

$A^{(8,1,0)}_{(15,1,1)}$& $A^{(8,1,0,0)}_{(15,1,1)}$& $A^{(15,1,0)}_{(15,1,1)}$&  $45_H$& $\frac{v^2_{45}}{\Lambda}$ \\ \hline
$A^{\prime(8,1,0)}_{(15,1,1)}$& $A^{\prime(8,1,0,0)}_{(15,1,1)}$& $A^{\prime(15,1,0)}_{(15,1,1)}$&  $45^\prime_H$& $\frac{v^2_{\mathrm{GUT}}}{\Lambda}$ \\ \hline
$A^{(1,3,0)}_{(1,3,1)}$& $A^{(1,3,0,0)}_{(1,3,1)}$& $A^{(1,3,0)}_{(1,3,1)}$&  $45_H$& $\frac{v^2_{\mathrm{GUT}}}{\Lambda}$ \\ \hline
$A^{\prime(1,3,0)}_{(1,3,1)}$& $A^{\prime(1,3,0,0)}_{(1,3,1)}$& $A^{\prime(1,3,0)}_{(1,3,1)}$&  $45^\prime_H$& $\frac{v^2_{\mathrm{GUT}}}{\Lambda}$ \\ \hline
$A^{\prime(3,2,-\frac{5}{6})}_{(6,2,2)}$ & $A^{\prime(3,2,-\frac{1}{2},-\frac{1}{3})}_{(6,2,2)}$ & $A^{\prime(6,2,\frac{1}{2})}_{(6,2,2)}$  & $45_H$ & $\frac{v_\mathrm{GUT}^2}{\Lambda}$ \\ \hline

$A^{(1,1,1)}_{(1,1,3)}$& $A^{(1,1,1,0)}_{(1,1,3)}$& $A^{(1,1,1)}_{(1,1,3)}$&  $45_H$& max$\lbrace v_{16},\frac{v_\mathrm{GUT}^2}{\Lambda}\rbrace$ \\ \hline
$\chi^{\prime(1,1,1)}_{(\overline{4},1,2)}$& $\chi^{\prime(1,1,\frac{1}{2},\frac{1}{2})}_{(\overline{4},1,2)}$& $\chi^{\prime(\overline{4},1,\frac{1}{2})}_{(\overline{4},1,2)}$&  $16_H^\prime$& min$\lbrace v_{16},\frac{v_{16}^2 \Lambda}{v_\mathrm{GUT}^2}\rbrace$ \\ \hline

$A^{\prime(\overline 3,1,-\frac{2}{3})}_{(15,1,0)}$& $A^{\prime(\overline 3,1,0,-\frac{2}{3})}_{(15,1,0)}$& $A^{\prime(15,1,0)}_{(15,1,1)}$&  $45_H^\prime$& max$\lbrace v_{16},\frac{v_\mathrm{GUT}^2}{\Lambda}\rbrace$ \\ \hline
$\chi^{\prime\prime(\overline 3,1,-\frac{2}{3})}_{(\overline{4},1,2)}$& $\chi^{\prime\prime(\overline 3,1,-\frac{1}{2},-\frac{1}{6})}_{(\overline{4},1,2)}$& $\chi^{\prime\prime(\overline 4,1,-\frac{1}{2})}_{(\overline{4},1,2)}$&  $16_H^{\prime\prime}$& min$\lbrace v_{16},\frac{v^2_{16}\Lambda}{v_\mathrm{GUT}^2}\rbrace$ \\ \hline
$\chi^{(\overline 3,1,-\frac{2}{3})}_{(\overline{4},1,2)}$& $\chi^{(\overline 3,1,-\frac{1}{2},-\frac{1}{6})}_{(\overline{4},1,2)}$& $\chi^{(\overline 4,1,-\frac{1}{2})}_{(\overline{4},1,2)}$&  $16_H$& $v_{45}$ \\ \hline
$\chi^{\prime(\overline 3,1,-\frac{2}{3})}_{(\overline{4},1,2)}$& $\chi^{\prime(\overline 3,1,-\frac{1}{2},-\frac{1}{6})}_{(\overline{4},1,2)}$& $\chi^{\prime(\overline 4,1,-\frac{1}{2})}_{(\overline{4},1,2)}$&  $16_H^{\prime}$& $v_{45}$ \\ \hline

$A^{\prime(3,2,\frac{1}{6})}_{(6,2,2)}$& $A^{\prime(3,2,\frac{1}{2},-\frac{1}{3})}_{(6,2,2)}
$& $A^{\prime(3,2,2,-\frac{1}{3})}_{(6,2,2)}$&  $45_H^\prime$& $v_{16}$ \\ \hline
$\chi^{\prime(3,2,\frac{1}{6})}_{(4,2,1)}$& $\chi^{\prime(3,2,0,\frac{1}{6})}_{(4,2,1)}$& $\chi^{\prime(3,2,1,\frac{1}{6})}_{(4,2,1)}$&  $16^{\prime}_H$& $v_{16}$ \\ \hline

$H^{\prime(1,2,\frac{1}{2})}_{(1,2,2)}$&$H^{\prime(1,2,\frac{1}{2},0)}_{(1,2,2)}$& $H^{\prime(1,2,\frac{1}{2})}_{(1,2,2)}$&  $10_H^\prime$& $\frac{v^2_{\mathrm{GUT}}}{\Lambda}$ \\ \hline 
$\chi^{\prime\prime(1,2,-\frac{1}{2})}_{(4,2,1)}$&$\chi^{\prime\prime(1,2,0,-\frac{1}{2})}_{(4,2,1)}$& $\chi^{\prime\prime(4,2,0)}_{(4,2,1)}$&  $16_H^{\prime\prime}$& $\frac{M_{16}v_{45}}{v_\mathrm{GUT}}$ \\ \hline

$\chi^{\prime\prime(\overline{3},1,\frac{1}{3})}_{(\overline{4},1,2)}$&$\chi^{\prime\prime(\overline{3},1,\frac{1}{2},-\frac{1}{6})}_{(\overline{4},1,2)}$& $\chi^{\prime\prime(\overline{4},1,\frac{1}{2})}_{(\overline{4},1,2)}$ & $16_H^{\prime\prime}$& $\frac{M_{16}v_{45}}{v_\mathrm{GUT}}$ \\ \hline
$H^{(3,1,-\frac{1}{3})}_{(6,1,1)}$&$H^{(3,1,0,-\frac{1}{3})}_{(6,1,1)}$& $H^{(6,1,0)}_{(6,1,1)}$&  $10_H$& $v_{45}$ \\ \hline
$H^{\prime(3,1,-\frac{1}{3})}_{(6,1,1)}$&$H^{\prime(3,1,0,-\frac{1}{3})}_{(6,1,1)}$& $H^{\prime(6,1,0)}_{(6,1,1)}$&  $10_H^\prime$& $v_{45}$ \\ \hline

\end{tabular}

\caption{ Approximate mass scales and mass eigenstates of the multiplets for gauge coupling unification for the scenario (b). Only the states that reside below the GUT scale and relevant for unification are presented. The supermultiplets shown in the first three rows are relevant for symmetry breakings below the GUT scale. }\label{tab:mixed-b-massscales}
\end{table}

\begin{figure}[t]
    \centering
    \includegraphics[width=0.49\textwidth]{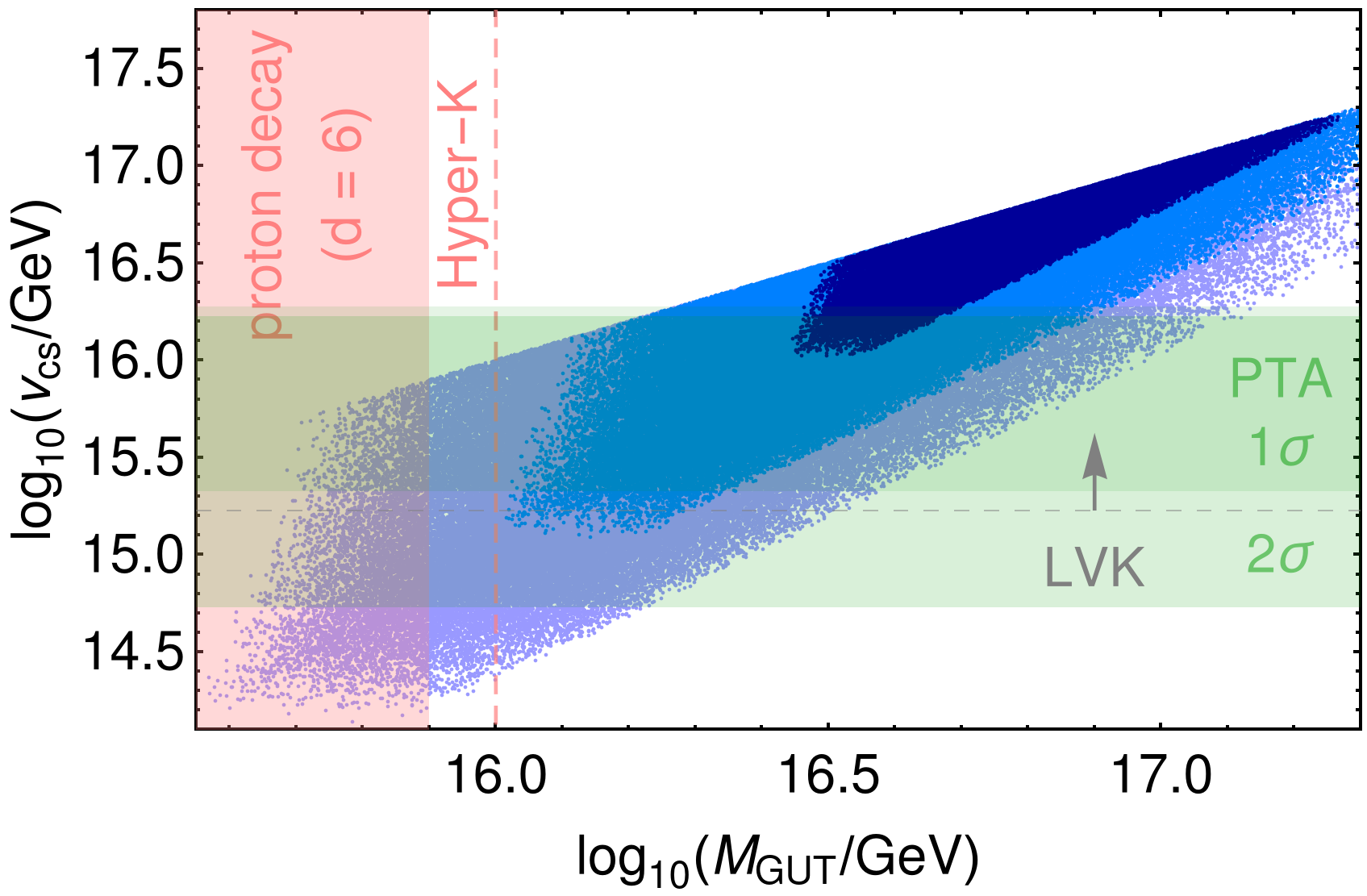}
    \hfill
   \includegraphics[width=0.49\textwidth]{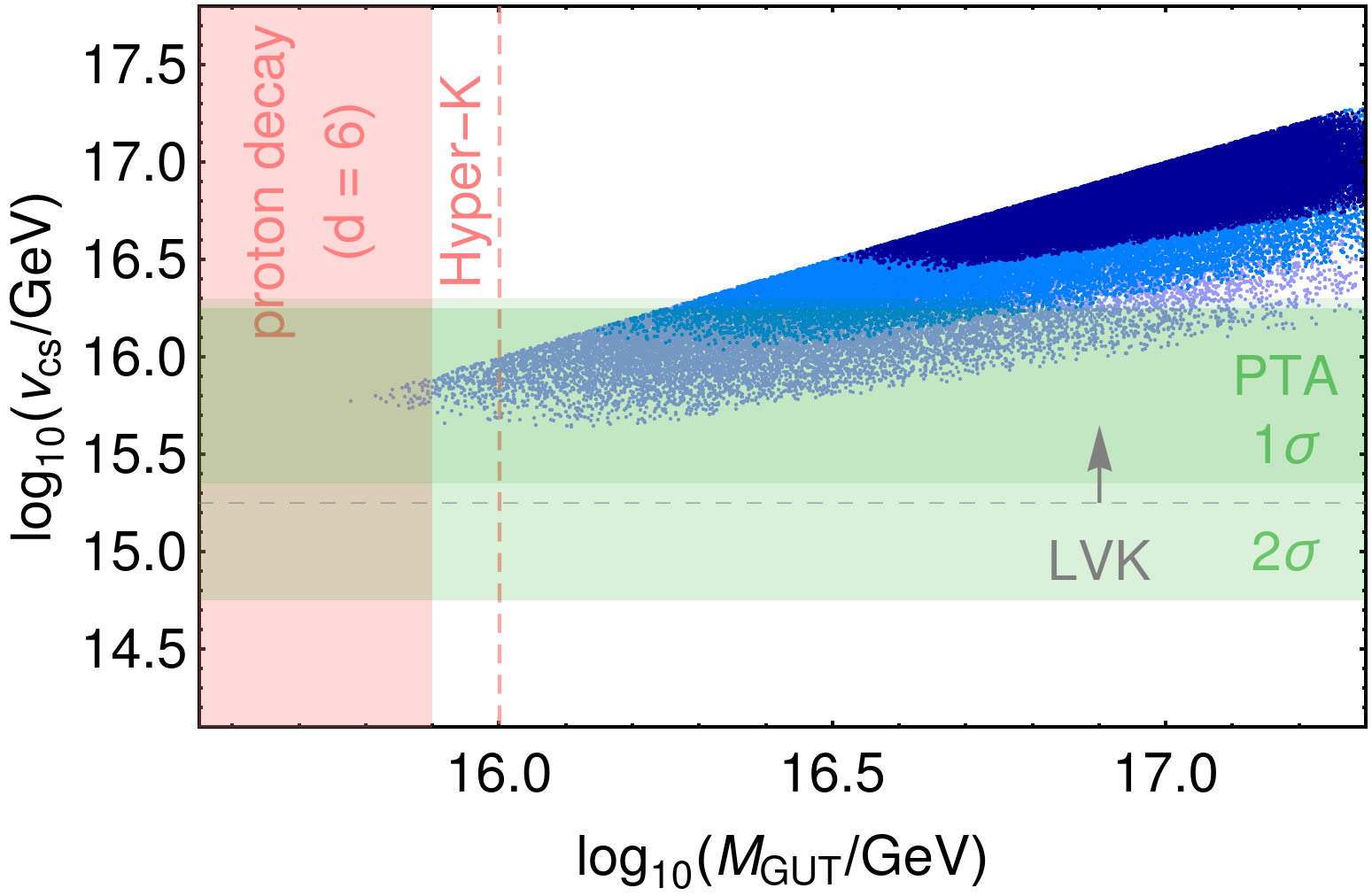}
    \caption{Preferred scales from gauge coupling unification for the two symmetry breaking chains: case (a) on the left panel and case (b) on the right panel. For dark/medium/light blue points the low-energy values of the gauge couplings $g_i$ agree with their measured values within $1\%$/$2\%$/$3\%$ uncertainties (cf.\ discussion on uncertainties in the main text). Assuming standard cosmological history, the current bound from the LVK collaboration is shown as dashed gray line, and the PTA preferred $68\%$ ($95\%$) credible interval on the string formation scale is shown as green (light green) area (see Sec.~\ref{sec:spectrum}). Here, $v_\mathrm{cs}=v_{16}$ represents the cosmic string formation scale.
    The current bound on the proton decay lifetime (from the dominant decay mode $p\rightarrow \pi^0e^+$) is presented as red shaded region, whereas the future sensitivity by Hyper-Kamiokande is shown as red dashed line. 
    Note that in case (c) we have $v_{16} \approx v_{45}$, which corresponds to the points at the upper boundary of both plots.
    }
    \label{fig:gcu_sigma3}
\end{figure}

Let us first consider the case (a), where $v_\mathrm{GUT}=v_{45}>v_{45^\prime}$. In this scenario, the $SO(10)$ gauge group is first broken to the left-right symmetric gauge group. The one-loop gauge coefficients in these two intermediate symmetry groups are
\begin{align}
    b^\textrm{3221}&=(b_{3C},b_{2L},b_{2R},b_{1B-L}^\prime)=(-3,1,4,15/2),\\
    b^\textrm{3211}&=(b_{3C},b_{2L},b_{1R},b_{1B-L}^\prime)=(-3,1,15/2,27/4),
\end{align} 
where the MSSM states along with the superfields responsible for breaking the intermediate symmetries are taken into account. If additional supermultiplets reside below a certain threshold, their contributions are added accordingly in the gauge coupling unification analysis.  For the $B-L$ factor, we have used the GUT normalization $b_{1B-L}^\prime=\frac{3}{2}\, b_{1B-L}$, and the hypercharge is matched with the other $U(1)$ factors by the relation 
\begin{align}
    \alpha^{-1}_Y(v_{16})=\frac{3}{5}\alpha^{-1}_{1R}(v_{16})+\frac{2}{5}\alpha^{\prime\,-1}_{1B-L}(v_{16}),\label{eq:hypercharge_matching}
\end{align}
while all other matchings are trivial.
Following the above-mentioned discussion, the masses of the intermediate-scale chiral multiplets are listed in Table~\ref{tab:mixed-a-massscales}. The possible points of gauge coupling unification (within $3\%$) are depicted in the left panel of Fig.~\ref{fig:gcu_sigma3}. From this figure, we find that the  GUT scale (which we identify with the heaviest gauge boson mass, cf.~Section~\ref{sec:mass matrices}) can take values within $10^{15.6}$~GeV and $10^{17.5}$~GeV, whereas $v_{16}$ can range from $10^{14.3}$~GeV up to $10^{17.5}$~GeV.

Now, in case (b), $v_\mathrm{GUT}=v_{45^\prime}>v_{45}$, i.e.~the scenario where the $SO(10)$ symmetry in the first step is broken to the quark-lepton symmetry, the one-loop gauge coefficients are given by 
\begin{align}
    &b^{\textrm{421}}=(b_{4C},b_{2L},b_{1R})=(-1,1,9),\\
    &b^{3211}=(b_{3C},b_{2L},b_{1R},b_{1B-L}^\prime)=(-2,1,15/2,27/4),
\end{align} 
where, as before, the MSSM states along with the superfields responsible for breaking the intermediate symmetries are taken into account. If additional supermultiplets reside below a certain threshold, their contributions to these coefficients are added accordingly in the study of the gauge coupling unification. 
For this second scenario, the mass scales of the additional chiral states are collected in Table~\ref{tab:mixed-b-massscales}. The obtained successful gauge coupling unification points within $3\%$ are shown in the right panel of Fig.~\ref{fig:gcu_sigma3}. This scenario corresponds to a GUT  scale ranges from $10^{15.9}$~GeV to $10^{17.5}$~GeV, while $v_{16}$ lies in between  $10^{15.6}$~GeV and $10^{{17.5}}$~GeV. We do not discuss case (c) separately, since it is obtained as limiting case with $\langle 45_H\rangle=\langle 45_H^\prime\rangle$.

\textbf{Gauge mediated proton decay:}  The exchange of the vector multiplets $X(3,2,-5/6)+Y(3,2,1/6)+c.c.$ leads to dimension six ($d=6$) proton decay. 
As mentioned above, since $d=5$ proton decay is suppressed by the way the DTS problem is resolved, the $d=6$ proton decay is expected to dominate. Among $d=6$ decay modes,
typically, the dominant decay channel is $p\rightarrow \pi^0e^+$, for which the current experimental bound on the partial lifetime is $2.4\times10^{34}$~years~\cite{Super-Kamiokande:2020wjk}. The expected lifetime of protons due to gauge-mediated processes is~\cite{Langacker:1980js}
\begin{align}
    \tau_p\sim \frac{16\pi^2 M_{X,Y}^4}{g_\mathrm{GUT}^4m_p^5},
\end{align}
where $m_p=938.33\,$MeV is the proton mass and $M_{X,Y}$ are the gauge boson masses (see Section~\ref{sec:mass matrices}). With $g_\mathrm{GUT}\sim0.9$ (which is a typical value in our gauge coupling unification analysis), we get a lower bound on the superheavy gauge boson masses of $M_{X,Y}\geq 10^{15.9}\,$GeV. Hyper-Kamiokande experiment will have the sensitivity to test the partial lifetime in the decay channel $p\rightarrow \pi^0e^+$ up to $7.8\times 10^{34}\,$years after 10 years of runtime~\cite{Hyper-Kamiokande:2018ofw}, which corresponds to gauge boson masses up to $ 10^{16.0}\,$GeV (for $g_\mathrm{GUT}\sim0.9$).

\section{Gravitational Wave Spectrum}\label{sec:spectrum}
As described in Sec.~\ref{sec:string}, in all  three breaking chains listed in Eqs.~\eqref{SB:a}-\eqref{SB:c}, we obtain hybrid topological defects, namely a network of metastable cosmic strings.  Due to the composite nature, this network eventually decays via the nucleation of monopole-antimonopole pairs along the string cores~\cite{Vilenkin:1982hm}.  The corresponding decay rate per string unit length is given by~\cite{Vilenkin:1982hm,Preskill:1992ck,Monin:2008mp,Leblond:2009fq,Chitose:2023dam}
\begin{align}
\Gamma_d\simeq \frac{\mu}{2\pi} e^{-\pi\kappa},\;\;\; 
\kappa= \frac{m^2}{\mu}\simeq \frac{8\pi}{g^2} \left( \frac{v_\mathrm{m}}{v_\mathrm{cs}} \right)^2.   \label{decay}
\end{align}
Here, $v_\mathrm{m}$  and $v_\mathrm{cs}$  are the VEVs associated with the  string and monopole formation scales, respectively. Furthermore, $\mu\simeq 2\pi v_\mathrm{cs}^2$ is the energy per unit length of the string and  $m\simeq 4\pi v_\mathrm{m}/g$ is the mass of the monopoles. Due to the exponential dependence in Eq.~\eqref{decay},  it follows that for squared monopole masses larger than the string scale such that $\kappa^{1/2}\gg 10$, the network can be considered effectively stable. 

For a metastable cosmic string network, the recent PTA results~\cite{NANOGrav:2023gor,Antoniadis:2023ott,Reardon:2023gzh,Xu:2023wog} hint towards a string tension in the range $G\mu \simeq \left(10^{-8} - 10^{-5}\right)$\footnote{Here, $G$ is the Newton's constant, which is defined as  $G=M^{-2}_\textrm{Pl}$ with $M_\textrm{Pl}=1.22\times 10^{19}$ GeV being the Planck mass.} and $\kappa^{1/2} \simeq \left(7.7 - 8.3\right)$, displaying a significant correlation between the two variables~\cite{NANOGrav:2023hvm}. For such a range of the metastability parameter $\kappa$,  the most stringent bound on the string tension comes from the LVK collaboration~\cite{LIGOScientific:2014pky,VIRGO:2014yos,KAGRA:2018plz}, which implies $G\mu \lesssim 2\times 10^{-7}$~\cite{LIGOScientific:2021nrg,NANOGrav:2023hvm}  at a frequency $\mathcal{O}(20)$ Hz~\cite{KAGRA:2021kbb}.    It is important to point out that assuming standard cosmology,  the $68\%$ credible regions in the $G\mu-\kappa^{1/2}$ parameter plane are in conflict with the aforementioned LVK bound. This is exciting since the LVK collaboration may very soon discover SGWB arising from the cosmic string network. 
 On the other hand, almost the whole $95\,\%$ credible interval is compatible with  the LVK O3
constraints. Intriguingly, from the gauge coupling unification analysis (Fig.~\ref{fig:gcu_sigma3}), one can see that the string formation scale is  compatible with the string tension preferred by the PTA data. The relation between these two quantities is approximately given by $G\mu\simeq 4.22\times 10^{-38} v^2_\mathrm{cs}$.

For the computation of the GW spectrum of a metastable cosmic string network, we follow the procedure described in Ref.~\cite{Antusch:2024ypp} (see also Refs.~\cite{Cui:2017ufi,Cui:2018rwi,Auclair:2019wcv,Gouttenoire:2019kij,Blasi:2020wpy,Dunsky:2021tih,Buchmuller:2021mbb}). Our scenario is supersymmetric, and following Sec.~\ref{sec:unification}, we fix $m_\mathrm{SUSY}=3$ TeV.   While presenting the results of the GW spectrum,   we consider two cases: (i) standard cosmology extended with additional SUSY degrees of freedom  and (ii)  a temporary matter domination phase in the evolution of the universe that somewhat dilutes the GW spectrum. As discussed in Ref.~\cite{Antusch:2024ypp}, the signs of the characteristic doubling of degrees of freedom predicted by SUSY could be probed at future GW detectors (for effects of new physics degrees of freedom on the GW spectrum of cosmic strings, see also Refs.~\cite{Battye:1997ji,Cui:2018rwi,Auclair:2019wcv}).

\begin{figure}[t]
    \centering
    \includegraphics[width=0.5\textwidth]{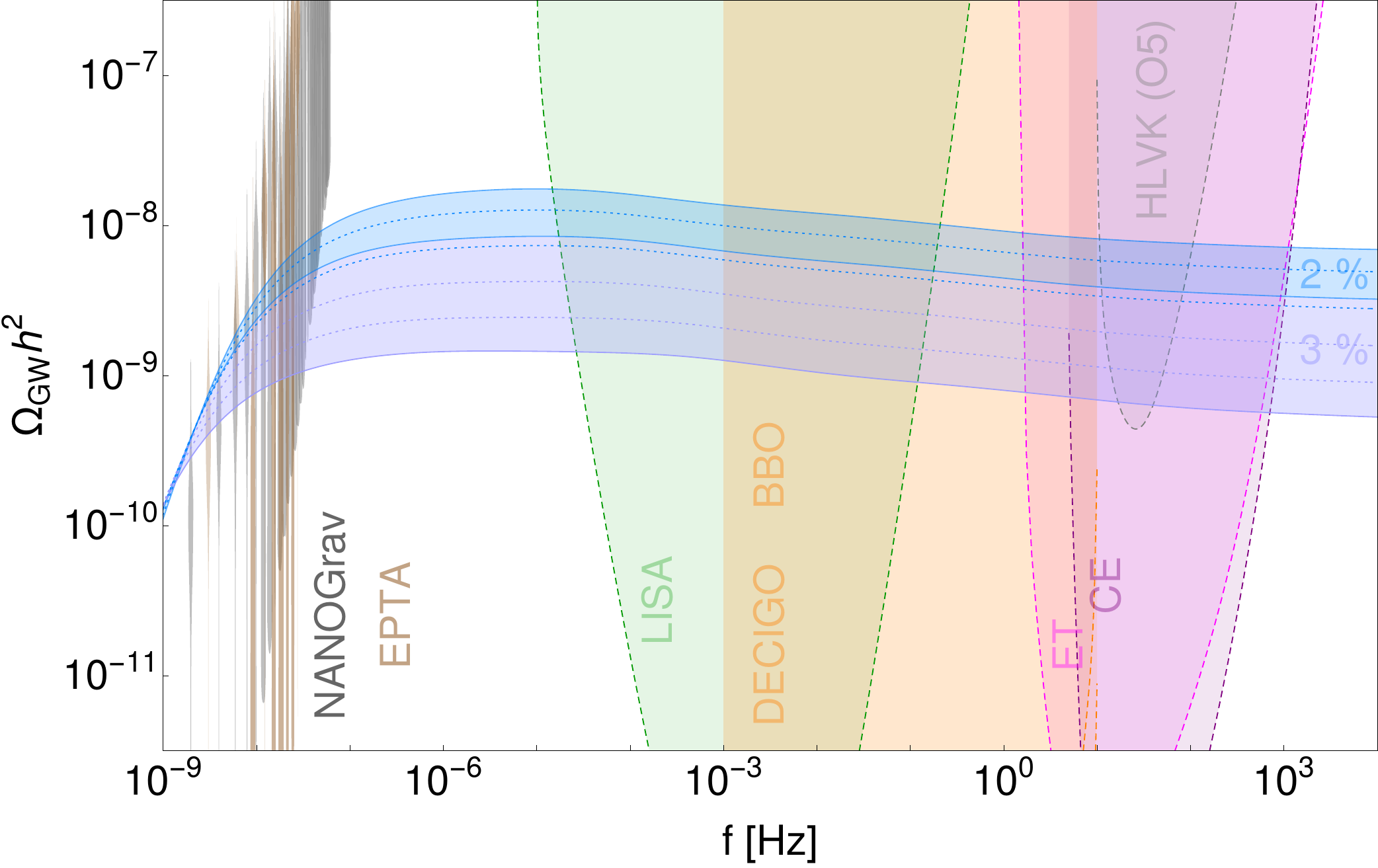}
    \caption{Gravitational wave spectrum of metastable cosmic strings assuming standard cosmology (and $M_\mathrm{SUSY}=3\,$TeV). For the case (a) breaking chain scenario, we show the part of the parameter space where gauge couplings unify (see Fig.~\ref{fig:gcu_sigma3}) within 2\% (medium) and 3\% (light) that is compatible with the current LVK O3 bound (with standard cosmology). Moreover, the dotted spectra correspond (from top to bottom) to choices of the parameters $(G\mu,\,\kappa^{1/2})$ of $(10^{-7},\,8.124)$, $(10^{-7.5},\,8.199)$, $(10^{-8},\,8.273)$, and $(10^{-8.5},\,8.348)$, respectively.}
    \label{fig:GW_undiluted}
\end{figure}

Beyond a standard cosmological history, we also consider the possible effects of late-time entropy production~\cite{Cui:2017ufi,Cui:2018rwi,Auclair:2019wcv,Gouttenoire:2019kij,Gouttenoire:2019rtn,Blasi:2020wpy,Ghoshal:2023sfa,Antusch:2024ypp}, caused e.g.\ by the late (but before BBN) decay of a modulus field. A possible candidate is the sgoldstino from a SUSY breaking hidden sector. We model it by an intermediate phase of matter domination (for details, cf.\ \cite{Antusch:2024ypp}), ending at redshift $z_E$ and leading to a certain dilution $\mathcal{D}$ of earlier produced GWs. In the GW spectrum from the metastable cosmic strings this causes a characteristic drastic drop of the GW spectrum beyond a certain frequency. This specific feature can be fully tested utilizing upcoming GW observatories. Owing to the suppression of the GW spectrum at high frequencies, early matter domination allows to raise the string tension  much above the currently allowed value by the LVK experiments, i.e., $G\mu\gtrsim 10^{-7}$. Therefore, with a certain amount of dilution, $\mathcal{D}\sim \mathcal{O}(10)$, a perfect fit to the latest PTA data that infers $G\mu\sim 10^{-6}$ is possible without violating the LVK bound.

The string tension, however, cannot be made arbitrarily high due to bounds from Big Bang nucleosynthesis (BBN)~\cite{Planck:2018vyg}, which successfully predicts the primordial abundances of light elements.  This prediction is sensitive to the effective number of degrees of freedom, $N_\mathrm{eff}$, at the time of the nucleosynthesis. Since the GW background contributes to the effective number of relativistic degrees of freedom, it is constrained by BBN bounds. Imposing the combined bounds from BBN and CMB, $\Delta N_\mathrm{eff}\lesssim 0.22$~\cite{Planck:2018vyg}, where  $\Delta N_\mathrm{eff}=N_\mathrm{eff}-N_\mathrm{eff}^\mathrm{SM}$ (with $N_\mathrm{eff}^\mathrm{SM}=3.044(1)$~\cite{EscuderoAbenza:2020cmq,Akita:2020szl,Froustey:2020mcq,Bennett:2020zkv}). This bound can be translated into a bound on the GW spectra~\cite{Maggiore:1999vm,Lazarides:2024niy} 
\begin{align}
\int^{f_\mathrm{high}}_{f_\mathrm{low}} \Omega_\mathrm{GW}(f) d\ln f \lesssim  5.7\times 10^{-6}\;  \Delta N_\mathrm{eff}\;.
\end{align}
For all spectra we show this bound is respected.

Let us now first consider the case within standard cosmology, including additional SUSY degrees of freedom. The result is shown in Fig.~\ref{fig:GW_undiluted}. In this plot, the series of gray and brown lines portray the recent SGWB observed in NANOGrav and EPTA, respectively. Moreover, the power-law-integrated sensitivity curves~\cite{Schmitz:2020syl} of various future gravitational wave observatories such as LISA, DECIGO, BBO, ET, and CE are illustrated with shaded color regions. As can be seen from Fig.~\ref{fig:gcu_sigma3}, the symmetry breaking chain (a) with intermediate left-right gauge group has the range  $10^{14.3}$ to $10^{17.5}$~GeV of $v_{16}$ that corresponds to a string tension in the range $G\mu\in(10^{-8.8},\,10^{-2.4})$.  The part of the parameter space where gauge couplings unify within $2\%$ and $3\%$ can be compatible with the current LVK bound, $G\mu\lesssim 2\times 10^{-7}$. The GW spectrum for the corresponding valid regions of the parameter space is depicted in Fig.~\ref{fig:GW_undiluted}. Within this band, the string tensions and the metastability parameters of a few benchmark spectra, highlighted with dotted lines, are also specified. 

It is evident from Fig.~\ref{fig:gcu_sigma3} that the symmetry breaking scenario (b) with intermediate quark-lepton gauge group (for this case, $v_{16}$  lies between $10^{15.6}$~GeV and $10^{{17.5}}$~GeV, which corresponds to  $G\mu\in(10^{-6.2},10^{-2.3})$ for the string tension) is disfavored when standard cosmology is assumed, because the LVK O3 bound conflicts with the entire $3\%$ regime preferred by gauge coupling unification. Therefore, if a SGWB is detected by the Advanced LVK’s (HLVK’s) upcoming observing run, O5, assuming standard cosmological history, scenario (a) would be preferred over scenario (b).

\begin{figure}[t]
    \centering
    \includegraphics[width=0.49\textwidth]{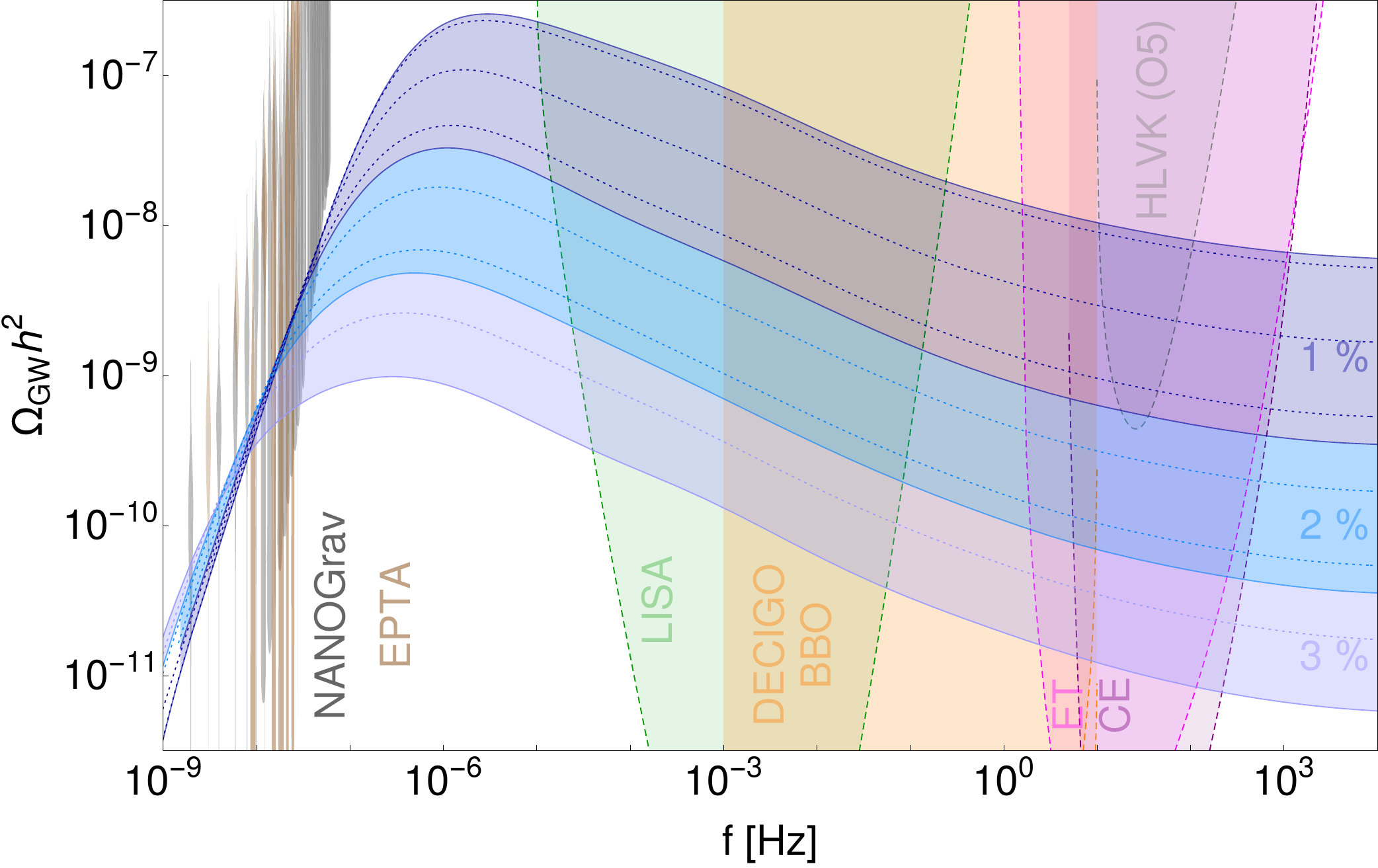}
    \hfill
    \includegraphics[width=0.49\textwidth]{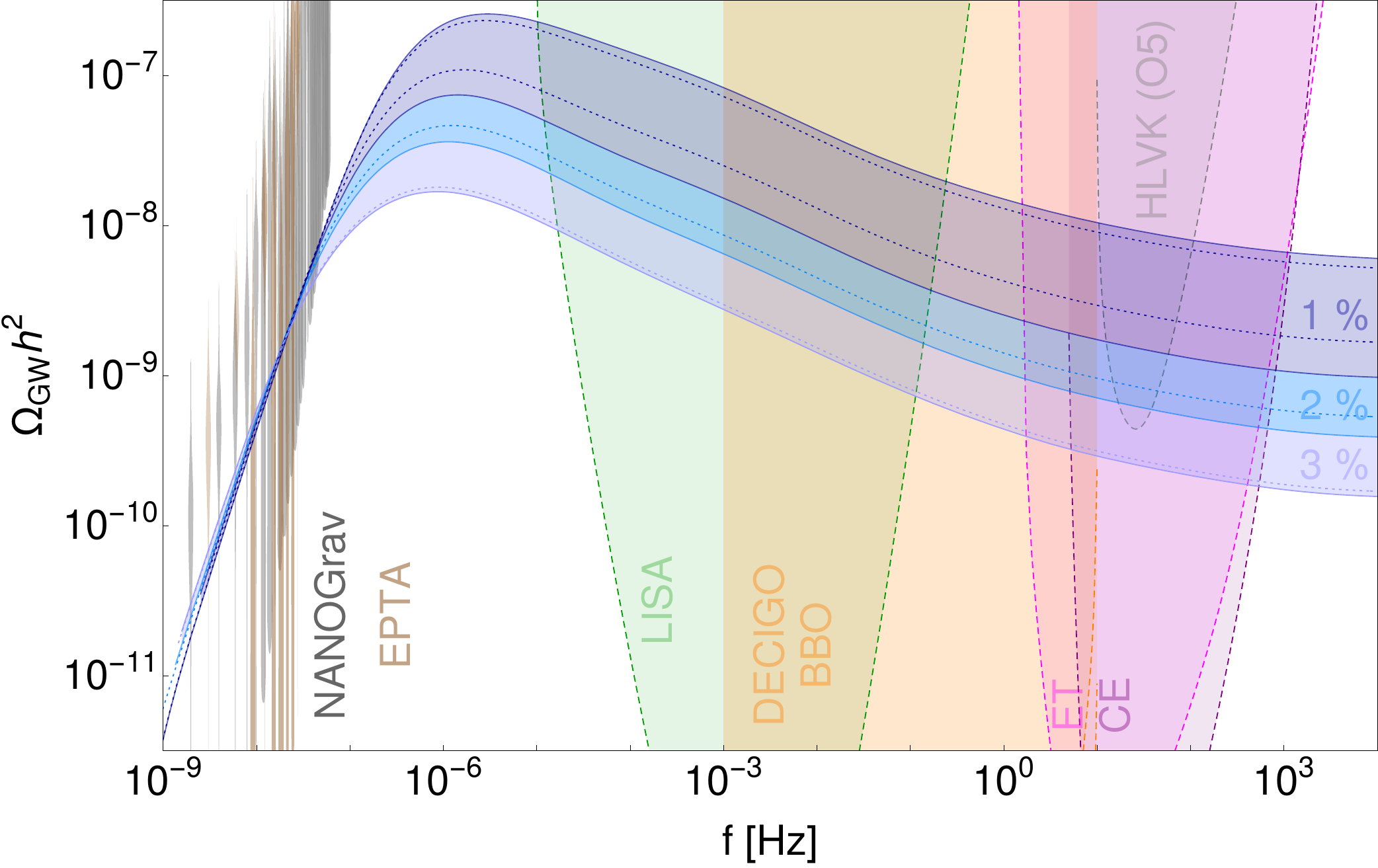}
    \caption{Gravitational wave spectrum of metastable cosmic strings with an additional phase of matter domination giving a dilution factor $\mathcal{D}=100$ (and $M_\mathrm{SUSY}=3\,$TeV). For the two breaking chain scenarios case (a) (left panel) and case (b) (right panel) we show the part of the parameter space, where gauge couplings unify (see Fig.~\ref{fig:gcu_sigma3}) within 1\% (dark), 2\% (medium), and 3\% (light). Moreover, in the left panel the dotted spectra correspond (from top to bottom) to choices of the parameters $(G\mu,\,\kappa^{1/2},z_E)$ of $(10^{-3},\,7.528,\,10^{12})$, $(10^{-4},\,7.677,\,10^{11.5})$, $(10^{-5},\,7.826,\,10^{11})$,  $(10^{-6},\,7.975,\,10^{10.5})$,  $(10^{-7},\,8.124,\,10^{10})$, and  $(10^{-8},\,8.273,\,10^{9.5})$, respectively. In the right panel the dotted spectra correspond to $G\mu=10^{-3}-10^{-6}$ with the same choices for $\kappa$ and $z_E$.   }
    \label{fig:GW_diluted}
\end{figure}

Next, we consider the effects of non-standard cosmology (more specifically, late time entropy production) on the GW spectrum, which may delay the discovery of the SGWB by LVK. The compatibility of gauge coupling unification within the $1\%$ range for scenario (a) and the entire $3\%$ range for scenario (b) suggests some amount of dilution of the GW spectrum during the early universe. As discussed above, we have modelled this as a temporary stage of late-time matter domination. The dilution from the late-time entropy production allows to raise the values of the string tension beyond the current (standard cosmology) LVK bound, i.e.\ it opens up the possibility for considering $G\mu\gtrsim 2\times 10^{-7}$. 

The GW spectra assuming a fixed dilution factor of $\mathcal{D}=100$ are presented in Fig.~\ref{fig:GW_diluted} for both the symmetry breaking channels (a) and (b). These plots show that the entire region where gauge coupling unification is preferably obtained can be made fully consistent with the current LVK bound with a certain amount of dilution. In Fig.~\ref{fig:GW_diluted}, although we have considered scenarios with large string tensions, we have checked that due to the dilution effects, the BBN bounds do not pose any constraint.  For clarity, we have explicitly provided the values of the string tension, the metastability parameter, and $z_E$ (the value of the redshift corresponding to the end of matter domination) for a series of lines, highlighted with dotted markings. If the GW spectrum of such a metastable cosmic string network is indeed discovered, then using data from future GW detectors like LISA, one would be able to reconstruct the dilution factor, which may provide fascinating model details.

Before concluding, we remark on the possibility of obtaining stable monopoles within our framework. Due to the exponential suppression when breaking up the strings, the only relevant monopole is the lightest intermediate one. For symmetry breaking chains (a) and (b), they correspond to the blue and red monopoles, respectively. In both scenarios, after eating up the string, the monopole and its corresponding antimonopole annihilate each other. In contrast, in symmetry breaking chain (c), all types of monopoles are produced at the same scale and are subsequently inflated away. The next symmetry breaking generates  strings that can additionally connect different types of monopoles, causing them to merge and resulting in the emergence of a stable monopole~\cite{Lazarides:2019xai}.  The present number density of such stable monopoles  can be within the observable range, provided that $G\mu$ lies between $10^{-9}$ and $10^{-5}$  for $\kappa^{1/2}\sim  4.55 - 5.53$~\cite{Lazarides:2024niy}. However, for such small values of $\kappa$, detection of the GW spectrum of metastable cosmic strings is challenging~\cite{Aggarwal:2020olq,Bringmann:2023gba,Servant:2023tua} since they exist only at high frequencies ($f\gtrsim 10^3$ Hz). Last but not least, the range of $\kappa^{1/2}$ indicated by the PTA data is much larger, which implies that the stable monopoles are not observables.

\section{Conclusions}\label{sec:conclusion}
Recently, pulsar timing arrays have detected tantalizing hints of a stochastic GW background at nanohertz frequencies. These signals could potentially arise from metastable cosmic strings, which are characterized by a GW spectrum that spans a vast range of frequencies. This distinctive feature paves the way for future GW observatories to test this hypothesis further. 

In this study, we have demonstrated how a network of such metastable cosmic strings can emerge within a promising supersymmetric $SO(10)$ grand unification framework. The proposed model class  seamlessly integrates cosmic inflation and addresses the doublet-triplet splitting problem without the need for fine-tuning, while also accounting for charged fermion masses and neutrino oscillation data. A particularly compelling aspect of the framework is that it predicts a unified scale that simultaneously accounts for inflation, neutrino mass generation, and the formation of the metastable cosmic string network. 

Our thorough analysis demonstrates that this unified scale is consistent with gauge coupling unification and complies with current experimental constraints on proton decay. We find that the scenario with an intermediate left-right symmetric group, case (a) is consistent with the current PTA data within standard cosmology, while the alternative scenario with an intermediate quark-lepton symmetric group, case (b), prefers a non-standard cosmological history with some dilution of the GW spectrum to be compatible with current LVK bounds. 

In summary, we have demonstrated that the ``promising'' model class singled out in Ref.~\cite{Antusch:2023zjk} indeed provides an attractive theory framework for explaining the PTA results via a network of metastable cosmic strings. Upcoming data from present and future gravitational wave observatories will test the metastable cosmic string hypothesis, has the potential to reveal the SUSY particle degrees of freedom (cf.\ Ref.~\cite{Antusch:2024ypp}) and can reveal a non-standard cosmic history if present in nature.

\section*{Acknowledgments}
We thank Jonathan Steiner for useful discussions.

\bibliographystyle{style}
\bibliography{reference}
\end{document}